\documentclass[12pt]{iopart}

\usepackage{graphicx,epsfig}
\usepackage[bookmarksnumbered,bookmarksopen]{hyperref}
\usepackage[capitalize]{cleveref}
\usepackage[utf8]{inputenc}

\AtBeginDocument{\usepackage{booktabs}}

\begin{document}

\title[Forced canonical thermalization]{Forced canonical thermalization in a hadronic transport approach at high density}

\author{Dmytro Oliinychenko$^{1,4}$ and Hannah Petersen$^{1,2,3}$\footnote{Emerging Leader}}

\address{$^1$ Frankfurt Institute for Advanced Studies, Ruth-Moufang-Straße 1, D-60438 Frankfurt am Main, Germany}
\address{$^2$ Institut f\"ur Theoretische Physik, Goethe-Universit\"at, D-60438 Frankfurt am Main, Germany}
\address{$^3$ GSI Helmholtzzentrum f\"ur Schwerionenforschung GmbH,  Planckstr. 1, 64291 Darmstadt, Germany}
\address{$^4$ Bogolyubov Institute for Theoretical Physics, Kiev 03680, Ukraine}
\ead{petersen@fias.uni-frankfurt.de}

\begin{abstract}
Hadronic transport approaches based on an effective solution of the relativistic Boltzmann equation are widely applied for the dynamical description of heavy ion reactions at low beam energies. At high densities, the assumption of binary interactions often used in hadronic transport approaches may not be applicable anymore. Therefore, we effectively simulate the high-density regime using the local forced canonical thermalization. This framework provides the opportunity to interpolate in a dynamical way between two different limits of kinetic theory: the dilute gas approximation and the ideal fluid case. This approach will be important for studies of the dynamical evolution of heavy ion collisions at low and intermediate energies as experimentally investigated at the beam energy scan program at RHIC, and in the future at FAIR and NICA. On the other hand, this new way of modelling hot and dense strongly-interacting matter might be relevant for small systems at high energies (LHC and RHIC) as well.
\end{abstract}

\pacs{1315, 9440T}
\vspace{2pc}
\noindent{\it Keywords}: relativistic hydrodynamics, hadron transport, hybrid approach, thermalization
\submitto{\jpg}
\maketitle
\normalsize


\section{Introduction}
\label{intro}

Relativistic heavy ion collisions have been performed during the last 30 years to study the properties of a strongly-interacting medium under extreme conditions and to explore the phase diagram of QCD (quantum chromodynamics) matter. Experiments encompass a broad range of energies from hundreds of MeV up to 5.5 TeV in the center of mass frame per nucleon pair and colliding systems from deuteron to uranium. One of the major goals of this research field is to study the transition from hadronic matter to the quark-gluon plasma (QGP). It is known from lattice QCD calculations that such a transition exists and that for matter at zero baryo-chemical potential, it is a smooth cross-over happening at a medium temperature of around 154 MeV \cite{Aoki:2006we,Soltz:2015ula}. Identifying the transition line and the search of the critical point is one of the major goals of heavy ion collision experiments nowadays, such as the beam energy scan program at RHIC (Relativistic Heavy Ion Collider), the NA61-SHINE experiment at CERN-SPS (Super Proton Synchrotron) and in the future at FAIR (Facility for Antiproton and Ion Research) and NICA (Nuclotron-based Ion Collider fAcility). Collisions of heavy ions accelerated almost to the speed of light are fast explosive processes lasting not more than 100 fm/c, so only the resulting debris are measurable in the detector. To interpret the results of the measurement and draw conclusions about the intermediate stages of collision, including the phase transition, theoretical effective dynamical descriptions such as the one presented here are required.

Relativistic fluid dynamics \cite{Huovinen:2001cy,Kolb:2003dz,Rischke:1995ir,Gale:2013da} and transport models based on the relativistic Boltzmann equation \cite{Molnar:2000jh,Xu:2004mz,Lin:2004en,Cassing:2009vt,Bass:1998ca,Bleicher:1999xi} are widely used to describe the dynamics of heavy ion collisions. The applicability ranges of these two approaches are complementary. Fluid dynamics can be applied, if the mean free path is significantly smaller than the system size and local equilibration is reached. Both of these conditions imply that the density is high enough with large collision rates to lead to local equilibration. Most of the transport models, on the contrary, rely on the low-density approximation, where one can neglect many-particle collisions. Whether a given part of the system is in the hydrodynamical regime or in the kinetic regime is defined by the Knudsen number - that is, the ratio of the mean free path to the system size, $Kn = \frac{l_{mfp}}{R}$ \cite{Niemi:2014wta}.

The complementarity of hydrodynamical and transport approaches makes hybrid approaches theoretically attractive, since in a hybrid approach each description is assumed to act in its region of applicability. Hybrid approaches are very successful at describing experimental data from the Relativistic Heavy Ion Collider (RHIC) and the Large Hadron Collider (LHC) \cite{Shen:2014vra, Werner:2010aa, Werner:2012xh, Petersen:2010cw, Petersen:2011sb, Petersen:2014yqa}. Currently, the focus in heavy ion collision experiments is shifting towards lower collision energies, at which the observation of the QCD critical point and the first order phase transition is expected. RHIC is performing a beam energy scan program, studying collision energies down to $\sqrt{s} = 2.5$ GeV; FAIR is under construction in Darmstadt and NICA in Dubna, aiming at studying low collision energies with high statistics. Theoretical simulations which explicitly involve an Equation of State (EoS), like hydrodynamics and hybrid approaches, are very important for these experiments because the EoS is one of the primary objects of their studies. At the same time, the application of hydrodynamics-based approaches at low energies meets challenges that were absent at high energies.

A typical hybrid approach starts with generating an initial state, which can be highly anisotropic and includes event-by-event fluctuations. Then a rapid switch to relativistic hydrodynamics is performed, which neglects the initial anisotropy of the energy-momentum tensor. Hydrodynamical equations are solved in the forward light-cone until some late time. The particlization hypersurface (usually a constant temperature, energy density or Knudsen number hypersurface) is then determined, a Cooper-Frye particlization \cite{Cooper:1974mv} is performed upon that surface, and particles are finally allowed to rescatter using hadronic transport, also called cascade or afterburner \cite{Huovinen:2012is}. Note that in such approaches, hydrodynamical equations are solved even out of their region of applicability, where the Knudsen number is large. The particlization hypersurface is determined a posteriori from hydrodynamics, but not from a dynamical condition considering both hydrodynamics and transport. Particles in the transport phase have no possibility to cause feedback to hydrodynamics, which leads to a well-known problem, the so-called negative Cooper-Frye contributions \cite{Bugaev:1996zq,Bugaev:1999wz,Anderlik:1998cb,Oliinychenko:2014tqa}. At high collision energies, at midrapidity, which is the kinematical region studied by RHIC and LHC, negative Cooper-Frye contributions are negligible and the approximation adopted by hybrid approaches is justified. At lower energies they can easily reach the level of 10\% for hydrodynamics with smooth initial conditions and are practically unlimited for event-by-event hydrodynamics \cite{Oliinychenko:2014tqa}.

Hydrodynamical and hybrid approaches could be completely substituted by transport models at low energies, but this presents two challenges. First, the EoS does not explicitly enter the transport model, so it becomes impossible to study the EoS directly, without specifying the degrees of freedom. Second, at high densities, multi-particle collisions gain importance. As an example, the account of $p\bar{p}$ annihilations to many mesons and the inverse process of many-meson collisions is claimed to be essential to describe anti-proton and anti-Lambda yields at AGS \cite{Cassing:2001ds}, as well as yields at the LHC \cite{Pan:2014caa}.

In this paper we explore a simple approach that attempts to solve or avoid the above mentioned problems. In a pure hadron transport model, we suggest to perform forced thermalization in the regions of high density. Physically, such thermalization corresponds to the extreme limit of N-particle collisions, so intense that thermalization happens rapidly, replacing the local distribution function by a thermal one. It follows from the H-theorem, that the thermalized state is unique and independent on the microscopic details of interaction, which makes it an easy case to consider. In fact such a treatment is conceptually very similar to a hybrid approach with Smoothed Particle Hydrodynamics \cite{Aguiar:2000hw}, but here hydrodynamics and transport are dynamically coupled. Forced thermalization involves the EoS, thus allowing to explore the phase transition. The method is also similar to core-corona separation \cite{Steinheimer:2011mp}, but the thermalized and transport domains are coupled dynamically and transport can feedback to the thermalized regions. It remains applicable for small systems and at low collision energies, where hydrodynamics or hybrid approaches are not applicable. All this motivates us to test and explore the implications of such an approach.

This paper is structured in the following way. In Section \ref{methodology}, the SMASH hadronic transport approach, which serves as a basis for implementing our approach, is presented and our methodology of the forced thermalization is explained. Different sampling algorithms are compared within a thermal box in Section \ref{box_test}. In Section \ref{results}, the behaviour of ideal hydrodynamics, transport and our approach within a simple expanding sphere setup are compared. Finally, at the end of that section, we apply our model to heavy ion collisions.

\section{Forced canonical thermalization}
\label{methodology}

As a framework for implementing effective N-particle scattering we employ the SMASH hadronic transport approach \cite{Weil:2016zrk} in the cascade mode neglecting nuclear potentials, Fermi motion and Pauli blocking. The most well-established hadrons and resonances listed in the Particle Data Group \cite{Agashe:2014kda} with existence rating marked as $^{***}$ or better are implemented. All particles are characterized by their 4-vectors in coordinate and momentum space and their masses (or spectral functions in the case of resonances). In cascade mode, particles can only propagate along straight lines or participate in $2 \leftrightarrow 1$ resonance formations/decays and $2 \to 2$ elastic/inelastic collisions. This limits our study to low collision energies, where string formation and fragmentation is negligible. In SMASH, particle production both in $pp$ and $AA$ collisions mainly happens via $NN \to N\Delta, \, \Delta\Delta, \, NN^*,\, NR,\, \Delta R$ and subsequent decays of $\Delta$ and $N^*$. Here $\Delta$ denotes all the excitations of the $\Delta(1232)$ resonance, $N^*$ denotes all the excitations of the nucleons, $N$ and $R$ denotes any resonance. Strangeness production is additionally affected by strangeness exchange reactions implemented following \cite{Graef:2014mra}. SMASH supports the test particle ansatz, where the number of sampled particles is scaled up with a factor $N_{test}$ and all the cross-sections are reduced by the same factor $N_{test}$, leaving the mean free path unchanged:
\begin{eqnarray}
N \to N \cdot N_{test} \\
\sigma_{ij} \to \sigma_{ij} N_{test}^{-1}
\end{eqnarray}
this test particle ansatz alleviates effects of non-locality in collisions \cite{Cheng:2001dz}; in the limit of $N_{test} \to \infty$, a cascade simulation effectively becomes a solution to the relativistic Boltzmann kinetic equation. In our study we take advantage of $N_{test} > 1$ to provide better statistics for the computation of local thermodynamic quantities, such as energy density, baryon density or pressure.


We implement the effective N-particle scatterings on top of the SMASH cascade approach as described above. The main assumption is that such scatterings happen, if the local rest frame energy density is high enough, and that they lead to a rapid thermalization. Indeed, the solution of the Boltzmann equation is a thermal distribution in the limit of N-particle scatterings and zero mean free path. In practice, the region $\Omega_{\epsilon_c}$ where the local rest frame energy density $\epsilon$ is larger than some predefined $\epsilon_c$ is determined and  particles in this region are substituted by new ones, sampled according to a thermal distribution conserving total energy, momentum and quantum numbers. In other words, we are replacing the non-equilibrium distribution function by a thermal one in the transport approach at energy densities $\epsilon > \epsilon_c$. This treatment is ideologically similar to hybrid (hydrodynamical + transport) models, see \cite{Becattini:2016xct} for example, but in our case the boundary between the ''hydrodynamical'' and transport regions is found dynamically and not aposteriori; also negative Cooper-Frye contributions are not emerging in our approach.

Technically, forced thermalization consists of two steps - coarse-graining to determine the macroscopic densities and thermodynamic properties and the sampling of the new particles. To coarse-grain a Cartesian grid is spanned over the region of interest. The number of cells in each direction is a parameter, but its variation in a reasonable range does not influence results as we show later (see Section \ref{results}). In each cell, the upper row of the energy-momentum tensor $T^{\mu 0}(r)$ and the baryon density $j_B^0(r)$ is calculated
\begin{eqnarray}
T^{\mu 0}(r) &=& \sum_i p^{\mu}_i K(\vec{r}-\vec{r_i}, \vec{p}) \\
j_B^0(r) &=& \sum_i B_i K(\vec{r}-\vec{r_i}, \vec{p}) \nonumber
\end{eqnarray}
Here $K(\vec{r})$ is a smearing factor used to smooth out statistical fluctuations of the density. Following \cite{Oliinychenko:2015lva}, we use
$K(\vec{r}) = (2\pi \sigma^2)^{-3/2} \gamma \, exp\left(-\frac{\vec{r}^2 + (\vec{r} \cdot \vec{u})^2}{2 \sigma^2} \right)$, where $\sigma$ is a smearing parameter, $\vec{u} = \gamma \vec{\beta} = \frac{\vec{p}}{m}$ and $\gamma = (1-\beta^2)^{-1/2}$. This smearing kernel behaves properly under Lorentz transformations. In all simulations, we take $\sigma = 1$ fm, except for the simulations with the sphere setup, where $\sigma = 0.3$ fm is used to avoid too much smearing and allow for a reasonable comparison of the results to hydrodynamics. 

Assuming that $T^{\mu\nu}$ and $j^{\mu}$ have the ideal fluid form $T^{\mu\nu}_{ideal} = (\epsilon + p)u^{\mu}u^{\nu} - p g^{\mu \nu}$, $j^{\mu}_{ideal} = n u^{\mu}$, and adding the equation of state (EoS) $p = p(\epsilon,n)$, we obtain the rest-frame quantities of interest. To this end,  the following system of equations is solved:
\begin{eqnarray}
    T^{00} = (\epsilon + p) \gamma^2 - p \\
    T^{0i} = (\epsilon + p) \gamma^2 \vec{v} \\
    j_B^0 = n \gamma \\
    p = p_{EoS}(n, \epsilon)
\end{eqnarray}

\begin{figure}
  \centering
  \includegraphics[width=0.5\textwidth]{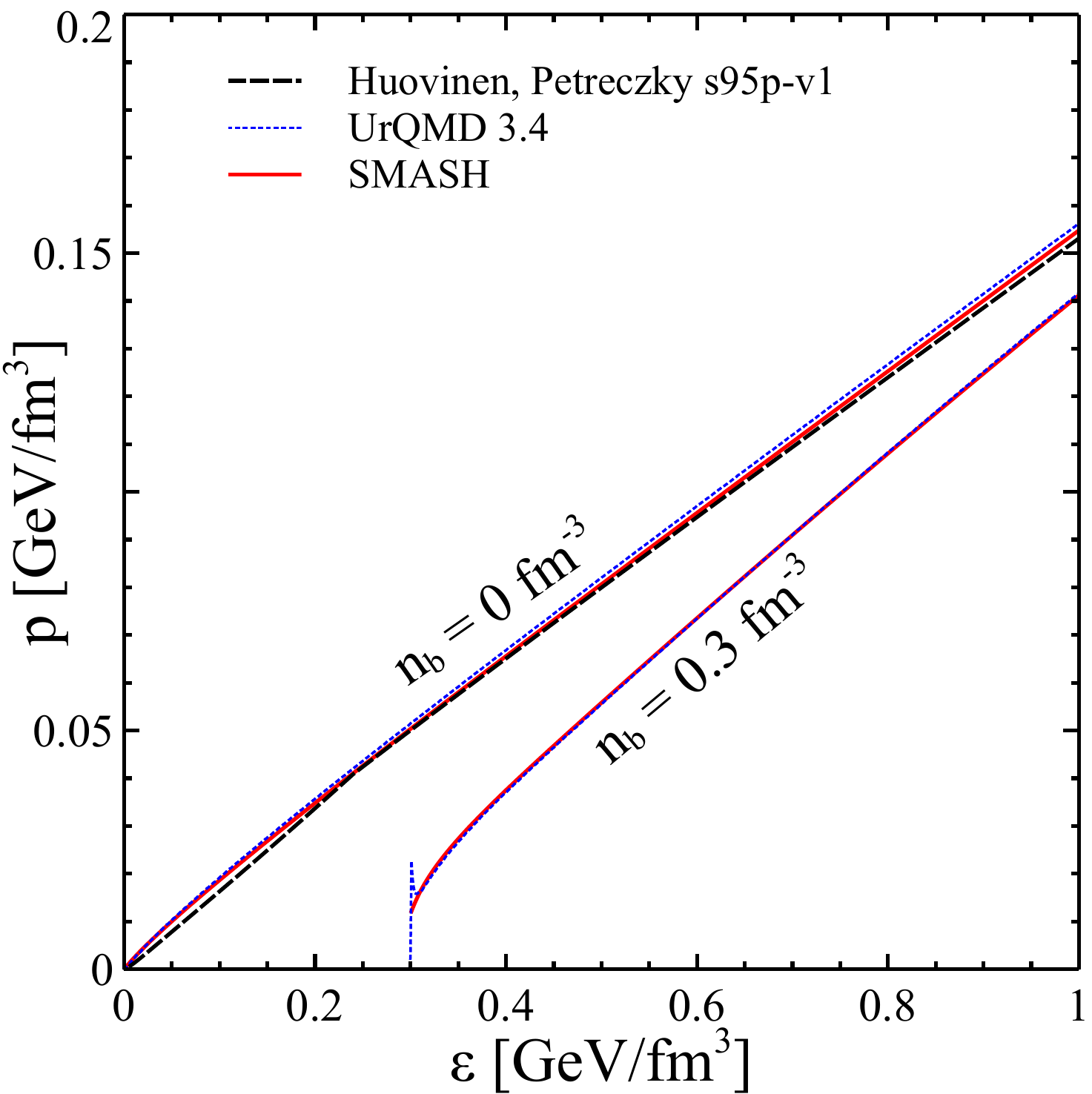}
  \caption{Equation of State (EoS) comparison between ideal gas consisting of hadrons implemented in SMASH (solid lines), in UrQMD (dotted lines) and s95p-v1 QCD EoS from \cite{Huovinen:2009yb} (dashed line).}
  \label{FIG:I}
\end{figure}

The procedure assumes that the strangeness density is negligible. For this procedure and also for extracting local temperatures and chemical potentials the EoS of an ideal Boltzmann hadron gas is employed consisting of all hadrons available in SMASH:
\begin{eqnarray}
  n_i &= \frac{T^3}{2\pi^2 (\hbar c)^3} g_i \lambda_i z_i^2 K_2(z_i) \\
  p &= T \sum n_i \\
  n_B &= \frac{T^3}{2\pi^2 (\hbar c)^3} \sum g_i B_i \lambda_i z_i^2 K_2(z_i) \\
  n_S &= \frac{T^3}{2\pi^2 (\hbar c)^3} \sum g_i S_i \lambda_i z_i^2 K_2(z_i) \\
  \epsilon &= \frac{T^4}{2\pi^2 (\hbar c)^3} \sum g_i z_i^2 \left(3K_2(z_i) + z_i K_1(z_i) \right) \,,
\end{eqnarray}
where the chemical potential $\mu_i \equiv \mu_B B_i + \mu_S S_i$ corresponds to baryon and strangeness conservation, $z_i \equiv \frac{m_i}{T}$ and fugacity $\lambda_i \equiv exp \left( \frac{\mu_i}{T} \right)$. This EoS does not take into account the effects of quantum statistics, consistently with the SMASH transport approach, where Pauli blocking is switched off and Bose enhancement is not implemented. The only small inconsistency between the above EoS and SMASH is that the width of resonances is not taken into account in the EoS. In Fig. \ref{FIG:I} the above EoS is compared with the hadron gas EoS from the UrQMD hybrid approach \cite{Petersen:2008dd} and the low-energy-density part of the QCD EoS from \cite{Huovinen:2009yb}, which is used in many hydrodynamical models. The difference between all three is negligible, which allows for a consistent comparison between our approach and the UrQMD hybrid approach.


After performing these steps, the information about the local rest-frame energy density $\epsilon(\vec{r})$, the temperature $T(\vec{r})$, the chemical potentials $\mu_B(\vec{r})$ and $\mu_S(\vec{r})$, and the local Landau rest frame velocity $v(\vec{r})$ is available in each cell of the grid. This allows to construct a region $\Omega_{\epsilon_c}$ where $\epsilon > \epsilon_c$, from which we remove particles and sample new ones according to the local $T$, $\mu_{B,S}$ and $v$.

Since this sampling procedure is not uniquely defined, let us now discuss a few possible options. In this discussion we denote the set of all conserved quantities (energy, momentum, baryon number, strangeness, electric charge, etc) in a given event in one cell by $C_{cell}$. The total conserved  quantities in the thermalization region in this event are $C_{tot} = \sum C_{cell}$, where the sum goes over all thermalization cells.

The first option is to apply the Cooper-Frye formula to every cell, as it is done at the particlization in many hydrodynamical models. Then the conservation laws are fulfilled in every cell, but only in the event average. In our notations, $\left\langle C_{cell}^{before} \right\rangle = \left\langle C_{cell}^{after} \right\rangle$, but $C_{cell}^{before} \neq C_{cell}^{after}$. Since we are working in the framework of a transport approach which strictly respects conservation laws in each event, it is desirable that the forced thermalization also follows conservation laws event-by-event.

Another option is to have exact event-by-event conservation laws, where $C_{tot}^{before} = C_{tot}^{after}$, but $C_{cell}^{before} \neq C_{cell}^{after}$ and $\left\langle C_{cell}^{before} \right\rangle \neq \left\langle C_{cell}^{after} \right\rangle$. This approach is applied for particlization in some hybrid models \cite{Petersen:2008dd, Huovinen:2012is}. We follow this idea, because it is reasonably fast and provides a very good approximation to the next approach, when it goes about the distribution of total (not cell by cell) hadron multiplicities. In the next section we investigate the implications of this choice and compare different algorithmic implementations of it. This makes our study useful for hybrid approaches, since we demonstrate how to perform particlization in a faster and more controlled way.

One more possibility is to perform microcanonical thermalization in each cell, so that $C_{cell}^{before} = C_{cell}^{after}$. This can in principle be done using the procedures described in \cite{Werner:1995mx} and \cite{Becattini:2004rq} for every cell. In this case, it seems that $T$ and $\mu$ are not necessary, but they are actually useful for initializing the Metropolis algorithm, as suggested in \cite{Becattini:2004rq}. This method has two disadvantages: first, Metropolis sampling is slow and the need to perform it in $\sim 10^4$ of cells makes it almost not feasible. The other disadvantage is a sensitivity to the cell size and $N_{test}$: indeed, in the case of very small cells there is typically one or zero particles in the cell. Resampling this one particle conserving all quantum numbers will most probably lead to no change at all. At the same time,  increasing $N_{test}$, one will find more than one particle in such a cell and thermalization results will change. So a combination of cell size and $N_{test}$ becomes a physical parameter, characterizing a radius of interaction.

As mentioned, in this argumentation we will use the second method. The initial hadrons are substituted by a new set of hadrons distributed with probability
\begin{eqnarray} \label{eq:prob_distr}
w(\vec{r_i},p_{i}) \sim \prod_{\mathrm{sorts}}\frac{1}{N_{sort}!} \prod_{i=1}^N \frac{d^3 \vec{r_i} d^3\vec{p_i}}{(2\pi\hbar c)^3} e^{-(p^{\nu}_i u_{\nu}(\vec{r_i}) + \mu_i(\vec{r_i}))/T(\vec{r_i})} \delta_E \delta_{\vec{p}} \delta_B \delta_S \delta_C
\end{eqnarray}

The $\delta$-functions in this expression denote the conservation of total energy, momentum, baryon number, strangeness and electric charge. All these quantum numbers should be equal to the quantum numbers of initial particles in the region $\Omega_{\epsilon_c}$. Without the $\delta$-functions, the sampling distribution from Eq. \ref{eq:prob_distr} is equivalent to a Cooper-Frye sampling on an isochronous hypersurface. In this case by integrating over $d^3 \vec{r_i} d^3\vec{p_i}$ one can easily see that the distribution of a particular hadron yield in one cell is Poissonian:
\begin{eqnarray} \label{eq:pois}
w(N_{i}) \sim \frac{(V_{cell}\varphi_i)^{N_i}}{N_i!} \\
\varphi_i = \frac{g_i e^{\mu/T}}{(2\pi\hbar c)^3}\int d^3p \, e^{-p^{\nu} u_{\nu}/T}
\end{eqnarray}
where $\varphi_i$ is the average equilibrium density of a given hadron species in the cell. Strictly speaking, in a case with total energy- and momentum conservation this consideration is not applicable any more, because now momentum integrations involve additional global $\delta$-functions, so the distribution in one cell may be different from Poissonian. However, we assume that there are many cells with many particles in them, so that the global conservation laws affect the local Poisson distributions only slightly. From these considerations it is clear, that the method prefers larger $N_{test}$ and not too large cells to achieve reliable results. The details of the sampling algorithm and a test in a thermal box are discussed in the next section \ref{box_test}.

Here we assumed Boltzmann statistics instead of more realistic Fermi-Dirac and Bose-Einstein statistics. This is done intentionally to be consistent with the absence of Pauli-blocking or Bose-enhancement effects in the transport simulation. For quantum statistics the multiplicity distribution in a cell is not Poissonian anymore. For bosons the mean multiplicity increases due to quantum statistics and the variance decreases, for fermions it is the opposite. We apply our model for low collision energies in the high-density region, where the typical temperatures are around 110 MeV and typical baryon chemical potentials are of order 700 MeV (see Fig. \ref{Fig:AuAu_Tmu}). The correction for pions is then $\frac{1}{2} \frac{K_2(2m_{\pi}/T)}{K_2(m_{\pi}/T)} \approx 7 \%$, for protons it is $\frac{1}{2} \frac{K_2(2m_p/T)}{K_2(m_p/T)} e^{\mu/T} \approx 4 \%$ and for all the other hadrons it has to be smaller.

We perform forced thermalization every $\Delta t_{th}$ starting from time $t_{start}$. Unless stated otherwise, we take $t_{start} = 0$ fm/c and $\Delta t_{th} = 1$ fm/c. Further we vary $\Delta t_{th}$ to see its effect on observables in Section \ref{results}. The system evolution before, between and after thermalizations follows the conventional SMASH cascade, with propagation, collisions and decays. We assume that the N-particle collisions happen momentarily, at a single point in time.

\section{Thermal box - testing the sampling algorithm}
\label{box_test}

\begin{figure}
  \includegraphics[width=\textwidth]{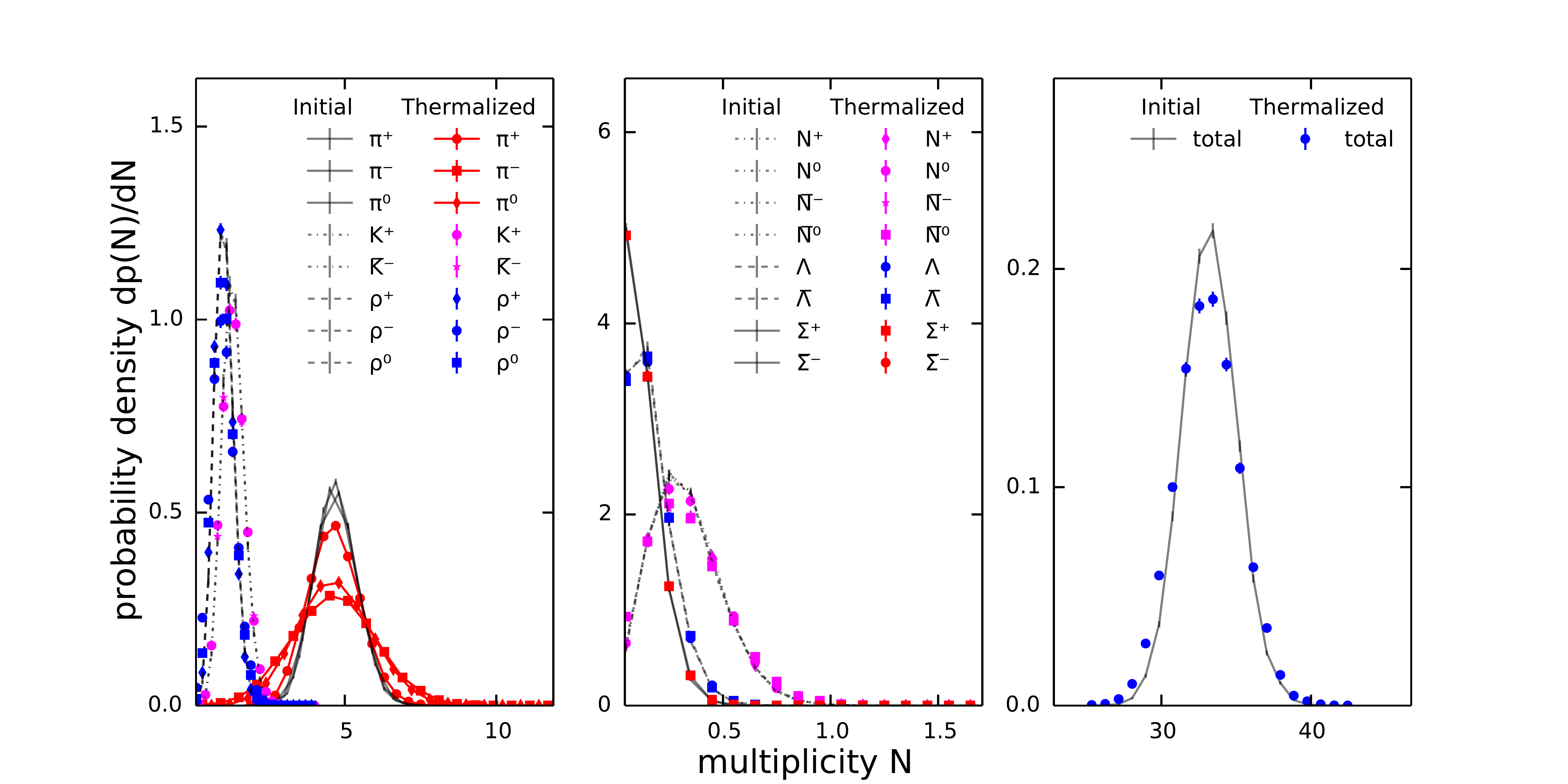}  
  \caption{Multiplicity distributions in a thermally initialized box are compared before and after additional forced thermalization. For a perfect thermalization algorithm these distributions should coincide. Here the thermalization algorithm is mode sampling.}
  \label{Fig:modes_sampling}
\end{figure}

Thermalization with global conservation of quantum numbers can be performed with different algorithms. In this Section, we compare three algorithms within a thermal box containing infinite hadronic matter in equilibrium. First, we create a $V = (5$ fm$)^3$ box with periodic boundary conditions and initialize thermally distributed hadrons that are available in SMASH. The multiplicities of each hadron species are $N_i = Poi(\phi_i)$, where $Poi$ is a Poisson distribution and $\phi_i$ is the thermal multiplicity of $i$-th hadron species at a temperature of $T = 0.15$ GeV and zero chemical potential $\mu_B = 0$. The values of temperature and chemical potential are an arbitrary choice, but they correspond to the relevant conditions in hybrid approaches for heavy ion reactions at high beam energies. The initial momenta are sampled from the Boltzmann distribution with the same temperature. The momenta in the box are centralized, so that total momentum of the box is zero, $p_i := p_i - \frac{1}{N}\sum_{j=1}^N p_j$. In this way we obtain a box with a thermalized hadron gas. The total energy and quantum numbers of such a box fluctuate from event to event.

As a second step the thermalization algorithm is applied, which conserves total energy, momentum and quantum numbers as described above in Section \ref{methodology}. The space-time grid consists of $10^3$ cells. On this grid, the coarse graining as described above is performed, taking the periodic boundary conditions into account. After sampling new particles with three different algorithms the initial multiplicity and momentum distributions with the ones after forced thermalization are compared. If everything works as expected, the results are supposed to be identical.

The first algorithm under investigation is the mode sampling used for particlization in the UrQMD hybrid approach \cite{Huovinen:2012is}. It consists of seven steps called ''modes'':
\begin{enumerate}
\item Choose a cell with probability $\frac{V_{cell}}{V}$. Sample particles in the cell according to the thermal distribution assuming a Poisson distribution around the mean, until the total energy exceeds $E_{init}$. Only particles containing $\bar{s}$ anti-quarks are kept, reject all the other particles.
\item Compensate strangeness by sampling only particles containing $s$-quarks.
\item Sample non-strange hadrons until the total energy exceeds $E_{init}$, keeping only non-strange baryons.
\item Compensate baryon charge by sampling only anti-baryons.
\item Sample non-strange mesons until the total energy exceeds $E_{init}$, keeping only positively charged non-strange mesons.
\item Compensate electric charge by sampling negatively-charged non-strange mesons.
\item Sample neutral mesons until the energy is conserved.
\end{enumerate}

In this manuscript, the original mode sampling algorithm has been improved to increase the computational speed. Choosing the cell with the  probability $\frac{N_{cell}}{\sum_{cells} N_{cell}}$ and sampling one particle definitely in there helps to avoid rejections and samples the same distribution in a faster way. This improvement is especially noticeable at high baryon chemical potential, such as the one reached in low energy heavy ion collisions. The average total number of anti-baryons can then be order of $10^{-5}$. Sometimes one or two anti-baryons are needed to compensate the baryon number, but the probability to sample one in the original algorithm is $10^{-5}$ divided by number of cells. Therefore, many rejection steps are avoided with the newly defined probability. 

Applying the mode sampling within the thermal box (see Fig. \ref{Fig:modes_sampling}), we observe that the mean values of multiplicities are all reproduced and many multiplicity distributions are also reproduced. However, the $\pi$ and $\rho$ multiplicity distributions are wider than the initial ones, which results in a wider distribution for the total multiplicity.  Moreover, the width of the multiplicity distribution follows $\Gamma(\pi^-) > \Gamma(\pi^0) > \Gamma(\pi^+)$, and similarly for $\rho$-mesons. To find the origin of this deviation from the expectation, the mode sampling order has been exchanged - instead of keeping only positively charged first and compensating with negative particles, we keep only negatively charged first and compensate with positive. After that we get $\Gamma(\pi^+) > \Gamma(\pi^0) > \Gamma(\pi^-)$. This demonstrates that the multiplicity distribution obtained from the mode sampling is sensitive to the internal algorithm realization, which is not physical. 

\begin{figure}
  \includegraphics[width=\textwidth]{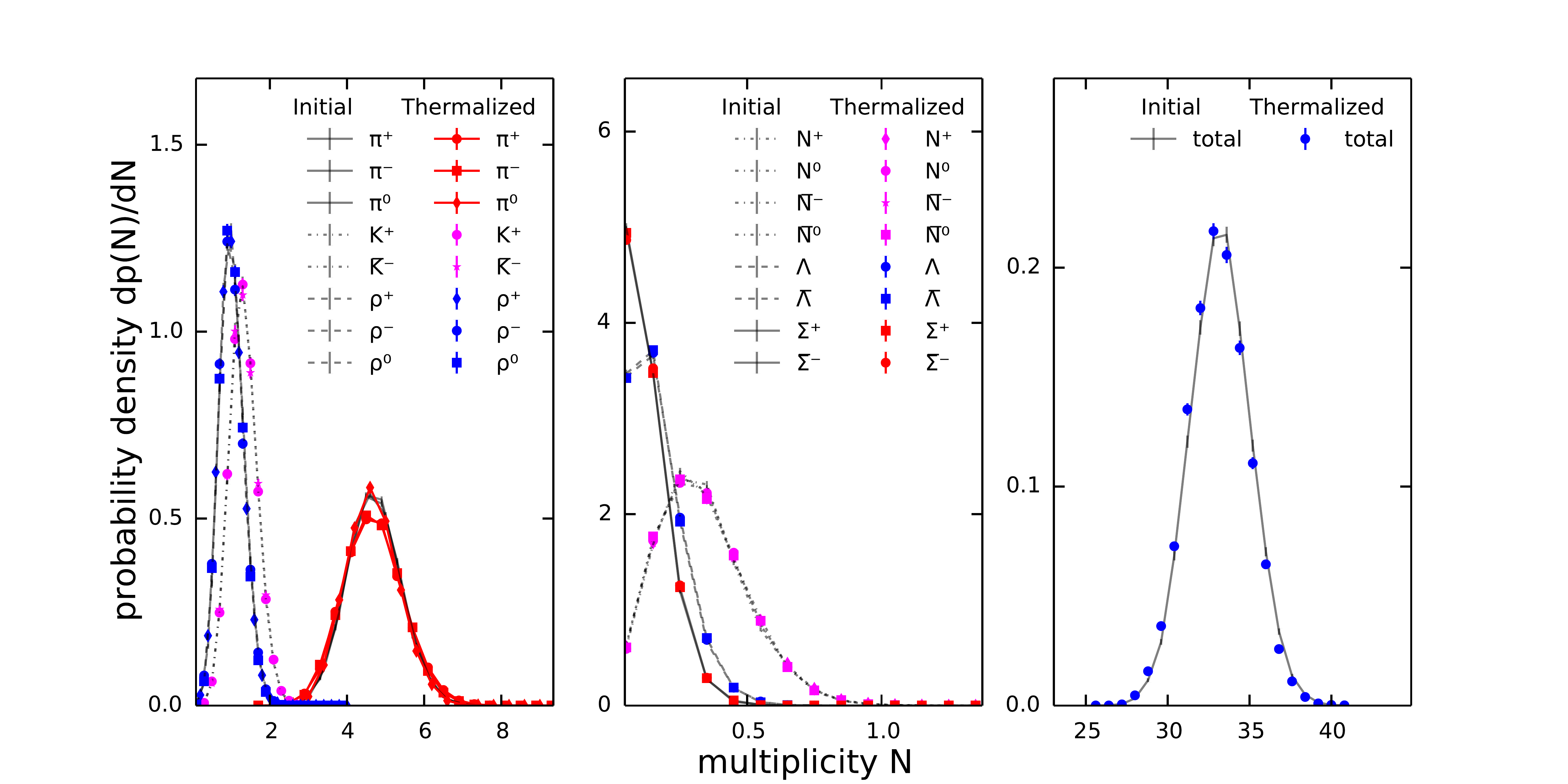}  
  \caption{Multiplicity distributions in a thermally initialized box are compared before and after additional forced thermalization. For a perfect thermalization algorithm these distributions should coincide. Here the thermalization algorithm is biased Becattini-Ferroni sampling with rejection by total energy.}
  \label{Fig:biasedBF_sampling}
\end{figure}

The second algorithm we consider is following the idea suggested by Becattini and Ferroni in \cite{Becattini:2004rq}, where one takes advantage of the fact that the sum of Poissonian variables is Poissonian itself. However, we implemented a biased version, different from the original Becattini and Ferroni suggestion. This version turns out to be numerically more efficient, while the bias is rather modest. The algorithm consists of the following steps:
\begin{enumerate}
\item Compute total average thermal numbers of baryons $\nu_{B}$ and anti-baryons $\nu_{\bar{B}}$. Sample $N_B$ and $N_{\bar{B}}$ with probability
\begin{equation}
w(N_B, N_{\bar{B}}) \sim \frac{\nu_{B}^{N_B}}{N_B!} \frac{\nu_{\bar{B}}^{N_{\bar{B}}}}{N_{\bar{B}}!} \delta(N_B - N_{\bar{B}} = B_{init}) \,.
\end{equation}
Such a distribution can be sampled very efficiently using the method discussed in the Appendix. Then the multiplicities of particular baryons and anti-baryons are sampled from the multinomial distribution.
\item Compute total thermal average  for strange and anti-strange mesons: $\nu_{S}$ and anti-baryons $\nu_{\bar{S}}$. Then sample $N_S$ and $N_{\bar{S}}$ with distribution
\begin{equation}
w(N_S, N_{\bar{S}}) \sim \frac{\nu_{S}^{N_S}}{N_S!} \frac{\nu_{\bar{S}}^{N_{\bar{S}}}}{N_{\bar{S}}!} \delta(N_S - N_{\bar{S}} = S_{init} - S_{\mathrm{sampled}}) \,.
\end{equation}
Then particular numbers of strange and anti-strange mesons are sampled from multinomial distribution.
\item Same procedure for charged non-strange mesons, in the distribution there is $\delta(N_C - N_{\bar{C}} = C_{init} - C_{\mathrm{sampled}})$, where $C_{\mathrm{sampled}}$ is the charge of the hadrons sampled in the previous steps.
\item Sample numbers of neutral mesons from Poissonian distributions.
\end{enumerate}

We notice that for this version of the algorithm the distribution of the total number of particles is too wide. It turns out that this effect can be decreased, if one rejects all samples where the energy is too far away from the initial energy. Rejecting $|E_{sampled} - E_{init}|/E_{init} > 1\%$, we obtain the correct distribution for the total multiplicity, but the sampled distributions for $\pi$ and $\rho$ are slightly wider than the initial ones, see Fig. \ref{Fig:biasedBF_sampling}. This algorithm is so efficient, because of the fast method to generate pairs of Poisson-distributed numbers with fixed difference, described in the Appendix. The simple rejection method used for this purpose in the original paper by Beccatini and Ferroni is fast enough for the case of small chemical potential, but becomes slow for $\mu_B \simeq 0.7$ GeV reached in the Au+Au collisions at $\sqrt{s} = 3$ GeV - the energy relevant for our investigation.

\begin{figure}
  \includegraphics[width=\textwidth]{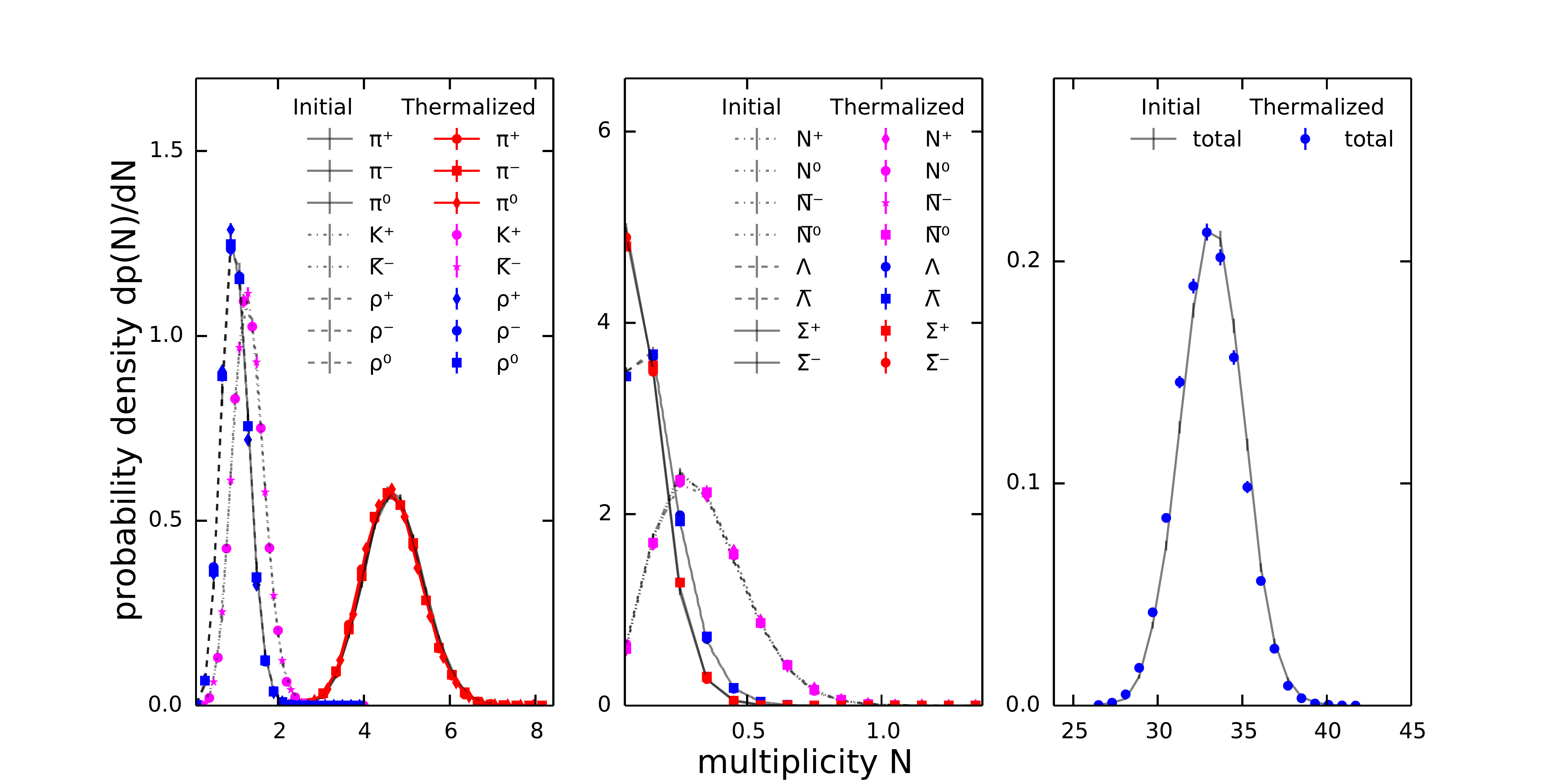}  
  \caption{Multiplicity distributions in a thermally initialized box are compared before and after additional forced thermalization. For a perfect thermalization algorithm these distributions should coincide. Here the thermalization algorithm is the unbiased Becattini-Ferroni sampling with rejection by total energy.}
  \label{Fig:BFefix_sampling}
\end{figure}

Finally, we test the unbiased algorithm, which is very similar to the previous one, except that rejection at any step requires the algorithm to start from scratch.

\begin{enumerate}
\item Identical to the first step of the biased algorithm.
\item Compute total thermal average for strange and anti-strange mesons: $\nu_{S}$ and anti-baryons $\nu_{\bar{S}}$. Then sample $N_S = Poi(\nu_{S})$ and $N_{\bar{S}} = Poi(\nu_{\bar{S}})$. If $N_S - N_{\bar{S}} = S_{init} - S_{\mathrm{sampled}}$, then proceed further, else start from the very beginning.
\item Similar to previous step for electric charge. If charge conservation not fulfilled, start from the very beginning.
\item Sample neutral non-strange mesons.
\item Sample momenta for all particles, if total energy deviates more than 1\% from the initial energy start from the very beginning.
\end{enumerate}

\begin{figure}
  \centering
  \includegraphics[width=0.5\textwidth]{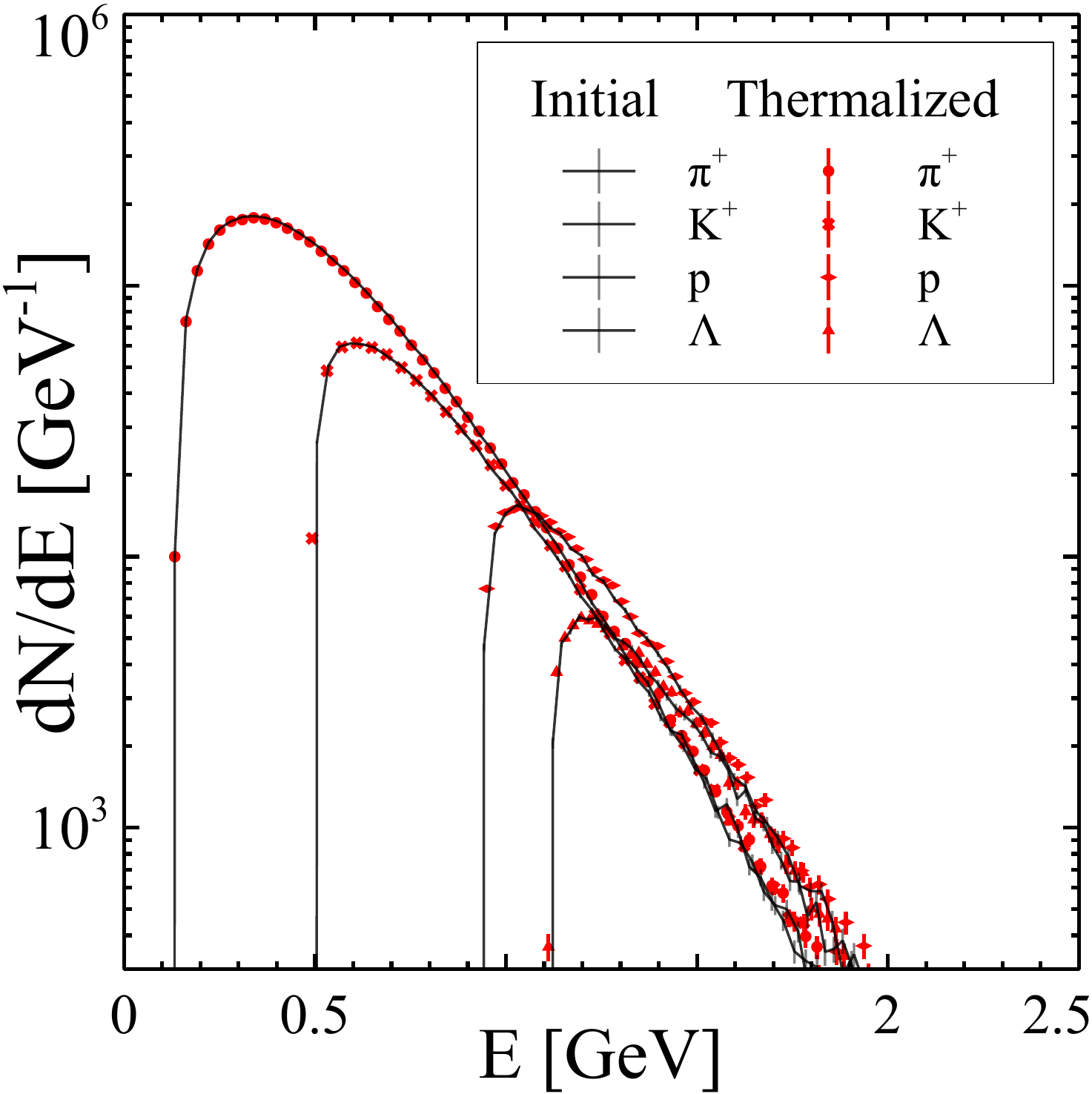}  
  \caption{Thermal box with the unbiased Becattini-Ferroni sampling with rejection by total energy: testing energy distribution}
  \label{Fig:BFefix_sampling_energy}
\end{figure}

This algorithm produces the correct multiplicity distributions (see Fig. \ref{Fig:BFefix_sampling}). Finally, Fig. \ref{Fig:BFefix_sampling_energy} shows that the energy distributions are also appropriate using this algorithm.

\begin{figure}
  \centering
  \includegraphics[width=0.5\textwidth]{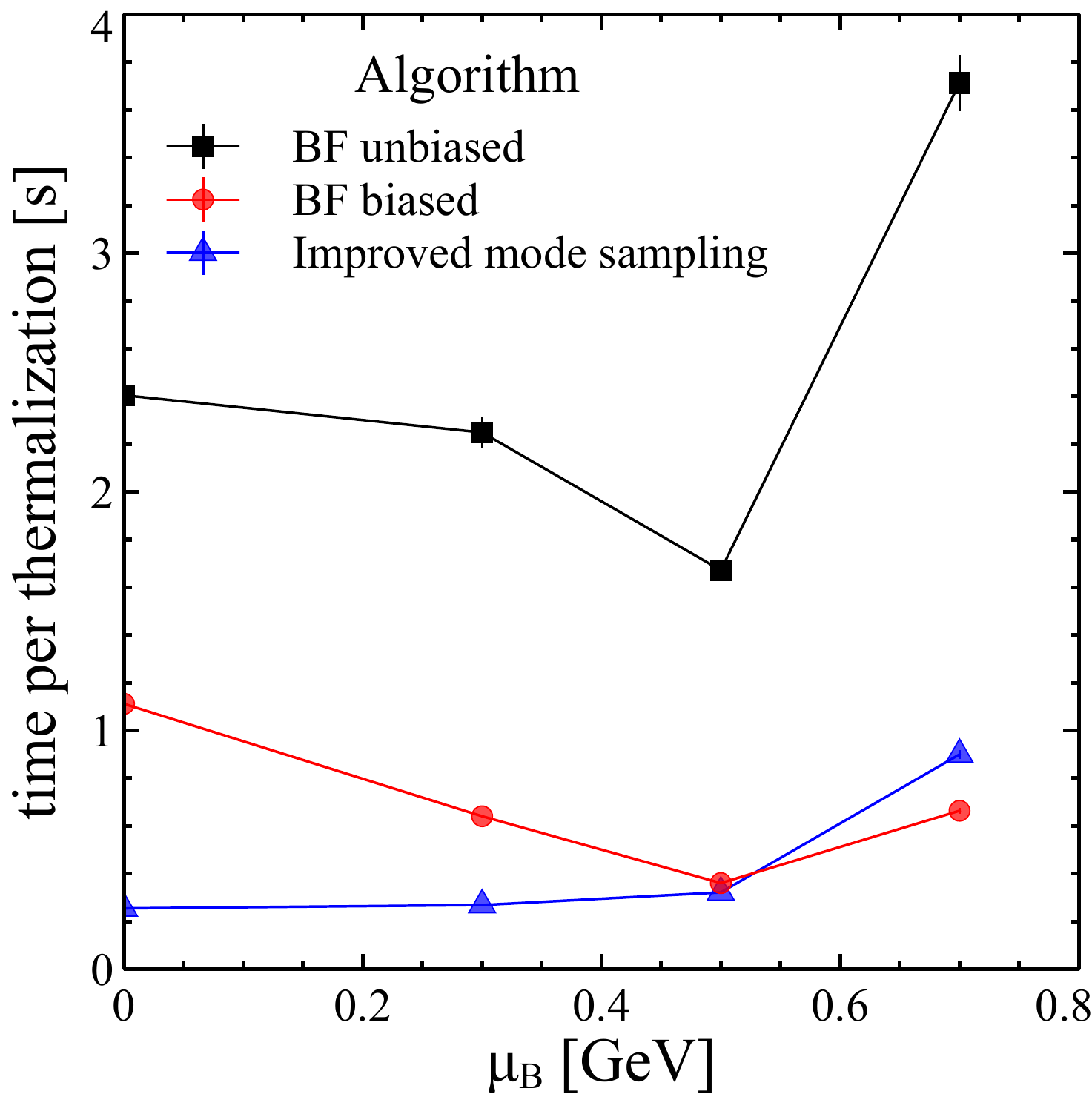}  
  \caption{Performance comparison of sampling algorithms: mode sampling, biased and unbiased Becattini-Ferroni sampling. Thermalization of the $(5$ fm$)^3$ thermal box with $T = 0.15$ MeV, $N_{test} = 10$, $\mu_S = 0$, $\mu_B$ is varied.}
  \label{Fig:algo_compar}
\end{figure}

In the following, the biased Becattini-Ferroni sampling is employed, since it is more efficient, the bias is small and it does not suffer from internal dependencies like the mode sampling. For a few cases, we also check that the unbiased algorithm produces identical results. In Fig. \ref{Fig:algo_compar} we compare the performance of the considered algorithms on an Intel(R) Xeon(R) 2.5 GHz CPU. For the summary of algorithm properties see Tab. \ref{Tab:algo_summary}.

After the application of the sampling algorithm quantum numbers are conserved, but energy is only conserved with 1\% precision and momentum conservation is only fulfilled on average. This shortcoming is addressed in two steps. First, we correct the momenta to match the initial momentum, $p_i := p_i - \frac{1}{N}(\sum_{j=1}^N p_j - p_{init})$. Then, we boost particles to the rest frame of initialization with 3-velocity $-\frac{p_{init}}{E_{init}}$. In this frame the sum of momenta is zero, because in the previous step we forced $\sum_{j=1}^N p_j = p_{init}$, so if one scales all momenta with the same factor, the sum will remain zero. Therefore, we scale all momenta with a factor $(1+a)$, such that
\begin{equation}
\sum_{j=1}^N \sqrt{m_j^2 + (1+a)^2 p_j^2} = E'_{init}
\end{equation}
We finally boost the particles back to computational frame and now energy and momentum are exactly conserved. This procedure biases momentum distributions, but this bias decreases with higher numbers of sampled particles $N$. One can observe in Fig. \ref{Fig:BFefix_sampling_energy}, that if $N$ is large enough, this bias is negligible.

\begin{table}
\caption{Sampling algorithms with total quantum numbers conservation}
\footnotesize\rm
\begin{tabular*}{\textwidth}{@{}l*{15}{@{\extracolsep{0pt plus12pt}}l}}
\br
Name & Bias on multiplicity distributions & Performance \\
\mr
Mode Sampling & \begin{tabular}{@{}l@{}} Widths of distributions affected,\\ bias dependent on implementation \end{tabular} & fast\\
Biased Becattini-Ferroni & \begin{tabular}{@{}l@{}}  Widths of distributions affected,\\ small bias observed \end{tabular} & comparable to Mode Sampling \\
Unbiased Becattini-Ferroni & No bias observed & $\simeq 4$ times slower than previous\\
\br
\end{tabular*}
\label{Tab:algo_summary}
\end{table}

\section{Interpolating between transport and hydrodynamics}
\label{results}

\begin{figure}
  \centering
  \includegraphics[width=0.49\textwidth]{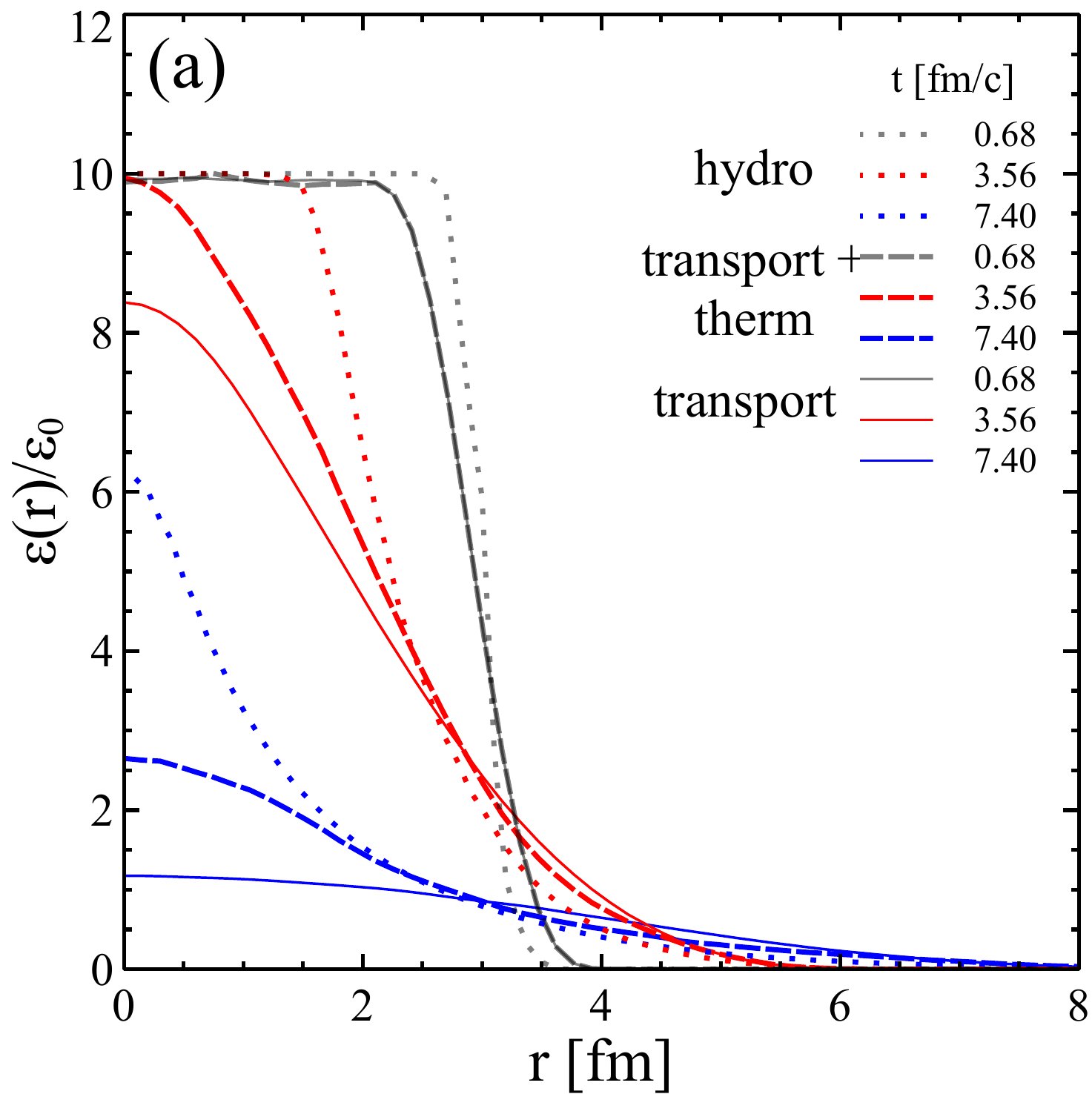}
  \includegraphics[width=0.49\textwidth]{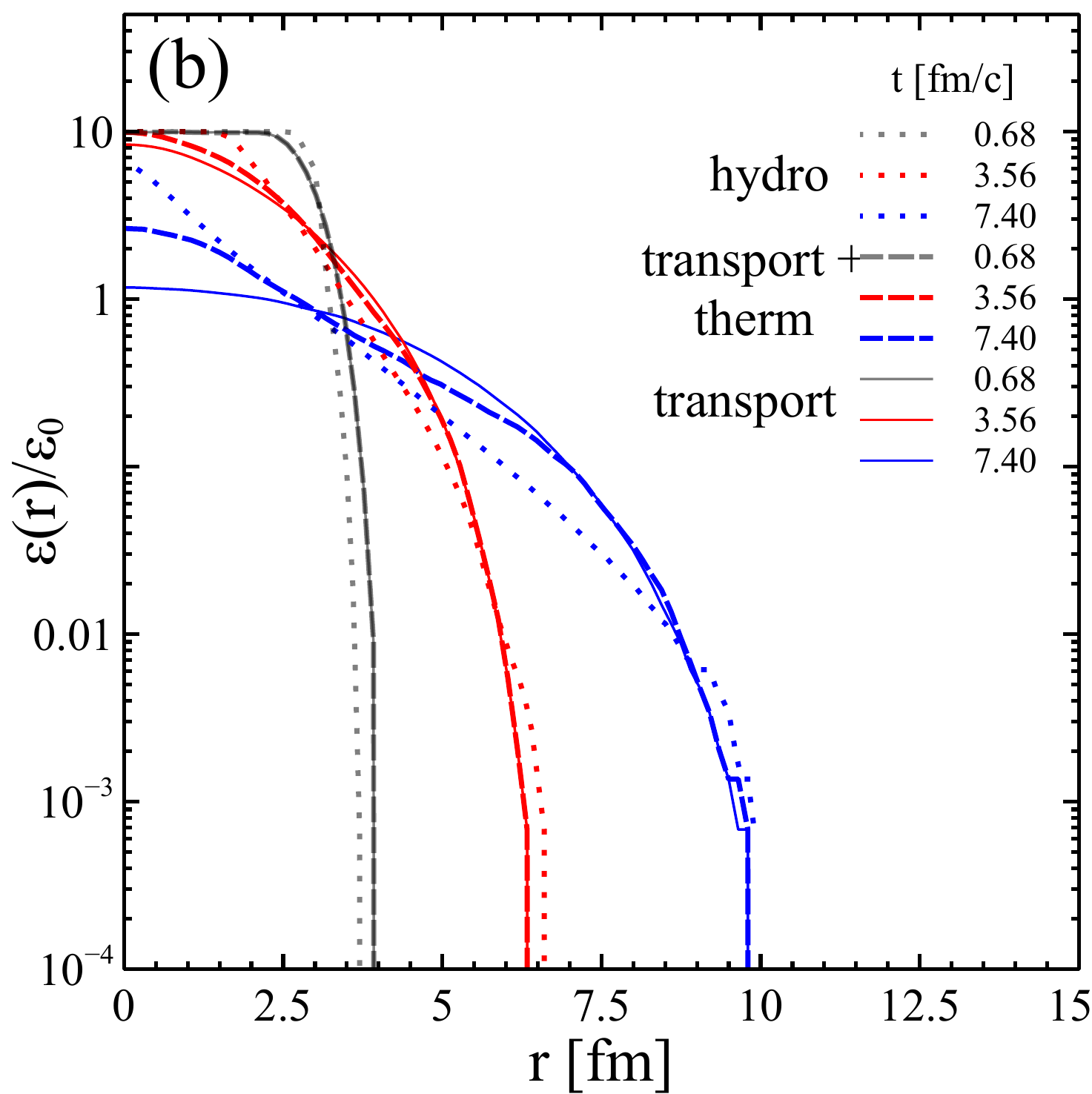} \\ 
  \includegraphics[width=0.49\textwidth]{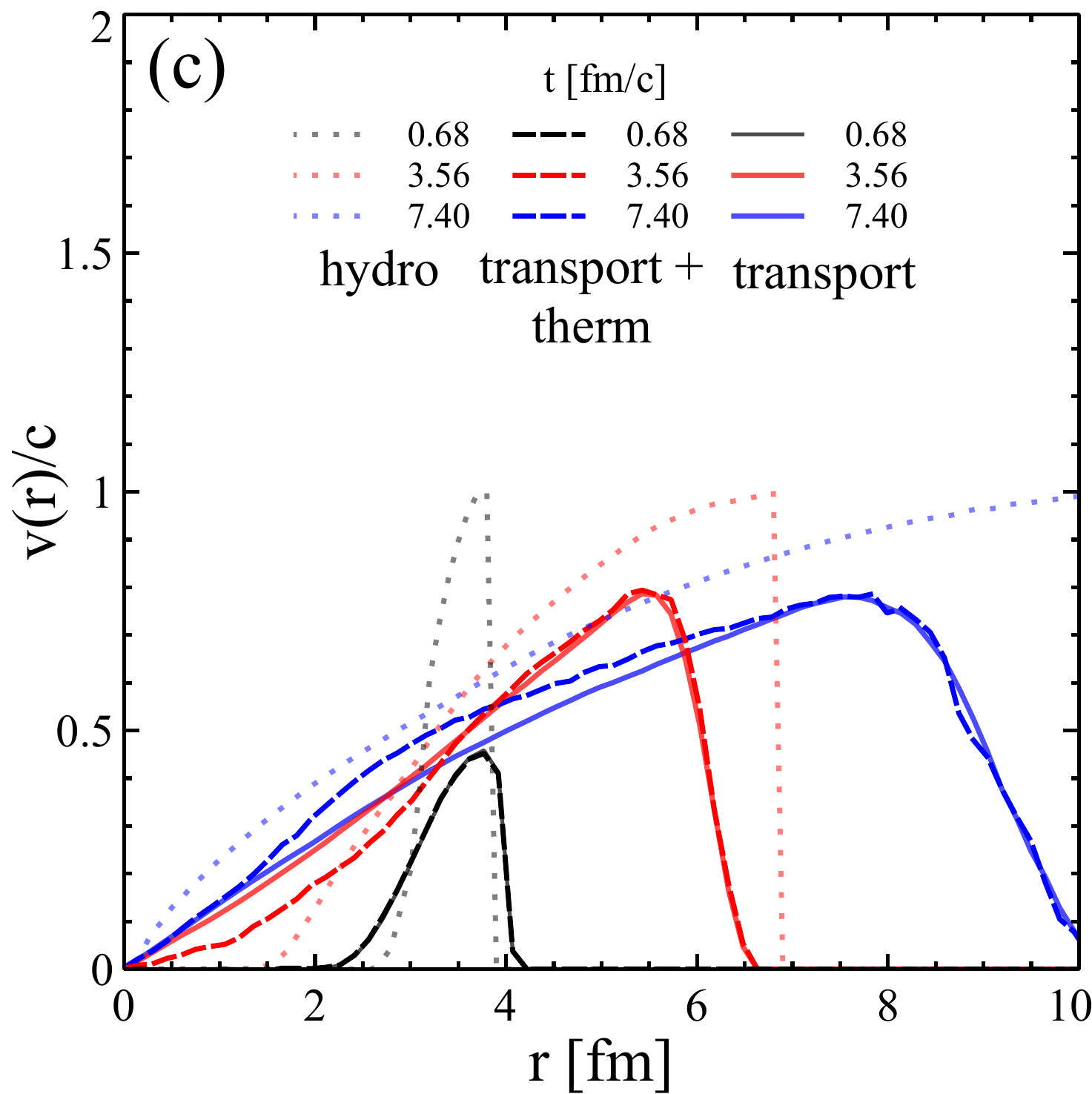}
  \caption{The time evolution of an expanding sphere is compared for ideal hydrodynamics (SHASTA, dotted lines), hadron cascade (SMASH, solid lines) and the same hadron cascade enhanced by effective N-particle collisions using forced thermalization (SMASH+therm, dashed lines). Panels (a) and (b) depict the energy density in the local Landau rest frame versus radius. Panel (b) is exactly the same plot as panel (a) with a logarithmic scale, which allows to see the edges of the system. Panel (c) demonstrates the velocity of the Landau rest frame versus radius. }
    \label{Fig:sphere}
\end{figure}

After establishing the details of the algorithm to effectively include N-particle collisions in a transport approach, we compare the influence on the time evolution of an expanding system to pure transport and ideal fluid dynamics. The idea is to prove that our original idea, that this dynamically coupled approach interpolates between the two limits of kinetic theory. For this purpose, a simple scenario is chosen, namely an expanding sphere. The sphere of radius $R_0 = 3$ fm is initialized at an energy density of 10 times nuclear ground state energy densities, $\epsilon = 10\epsilon_0$, and at zero baryon density. In Fig. \ref{Fig:sphere} the evolution of the local Landau rest frame energy density and velocity as a function of the radius $r$ are compared. The ideal hydrodynamics code has been performed using the SHASTA \cite{Rischke:1995ir} ideal fluid dynamics solver, which uses a $200^3$ Cartesian grid with $0.1$ fm grid spacing in each direction. The time step in SHASTA is 0.04 fm/c. In SMASH, the sphere is initialized with a thermal hadron gas with a temperature of $T \approx 191$ MeV, corresponding to energy density $\epsilon = 10\epsilon_0$. To minimize the effects of smoothing, we take the width of the Gaussian smearing kernel $\sigma = 0.3$ fm and compensate this by taking $N_{test}$ = 100. In the version of SMASH with the effect of N-particle collisions, forced thermalization is performed every $\Delta t_{th} = 1$ fm/c in the region, where energy density is above $2\epsilon_0$. The thermalization grid has a cell spacing of 0.5 fm, which can seem rather large, but we have checked that decreasing it by factors of 2 and 3 does not change the results. 

In Fig. \ref{Fig:sphere} one immediately notices that transport and fluid dynamics do not produce identical results, as expected. At the time when, in fluid dynamics, the rarefaction wave has still not reached the center, in transport the energy density at the center has already dropped. To understand this difference, one has to consider the Knudsen number in the transport case. At small times, the scattering rate in SMASH is 0.73 fm$^{-1}$, so that the mean free path is $l_{mfp} \simeq 1.5$ fm and $Kn \simeq \frac{l_{mfp}}{R_0} \simeq 0.5$. At this Knudsen number hydrodynamics is already on the verge of applicability. Moreover, this number is averaged over space and over various hadron species. On the edges of the system one has to compare the mean free path not to $R_0$, but to the distance to the edge. Furthermore, some hadron species have small interaction cross-sections with other particles, so their mean free path is large and they are in the ballistic regime, not in the hydrodynamic one. In Fig. \ref{Fig:sphere}, panel (c) shows that velocity at the edge in the hydrodynamics is $c$, while it is smaller in the transport model, because of massive particles being present.

SMASH including the effective treatment of N-particle collisions exhibits intermediate behaviour between hydrodynamics and pure transport. At the edges of the system, where forced thermalization is not happening, it behaves like transport, while in the center, it is closer to hydrodynamics. By forcing thermalization every $\Delta t_{th} = 1$ fm/c, the Knudsen number at the center is fixed for some time to $Kn \simeq \frac{\langle v_{therm} \rangle \Delta t_{th} }{R_0}$ for all hadron species. So even for hadrons with small cross-sections it becomes hard to escape the center too early. In fact, one can regulate this closeness to hydrodynamics by changing $\Delta t_{th}$. For smaller $\Delta t_{th}$ one obtains smaller Knudsen number and the result is thus closer to hydrodynamics. We underline that the region of forced thermalization is coupled to the outside region: particles can move in both directions. This is different from hybrid approaches, where particles from transport have no chance to feedback to hydrodynamics. Overall, the introduction of effective N-particle collisions has the expected effect that it interpolates between pure transport and ideal hydrodynamics. 

\begin{figure}
  \centering
  \includegraphics[width=0.49\textwidth]{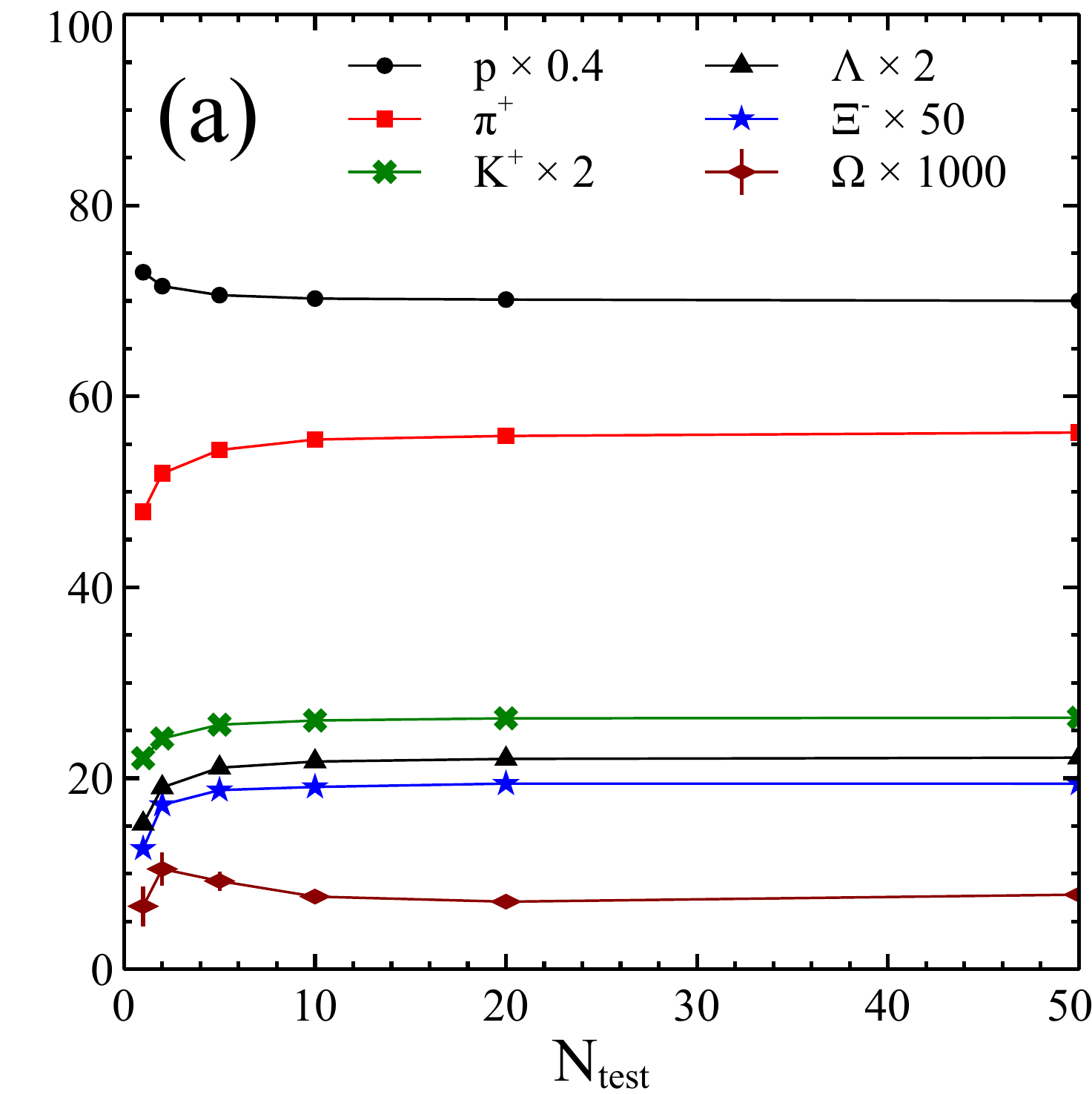}
  \includegraphics[width=0.49\textwidth]{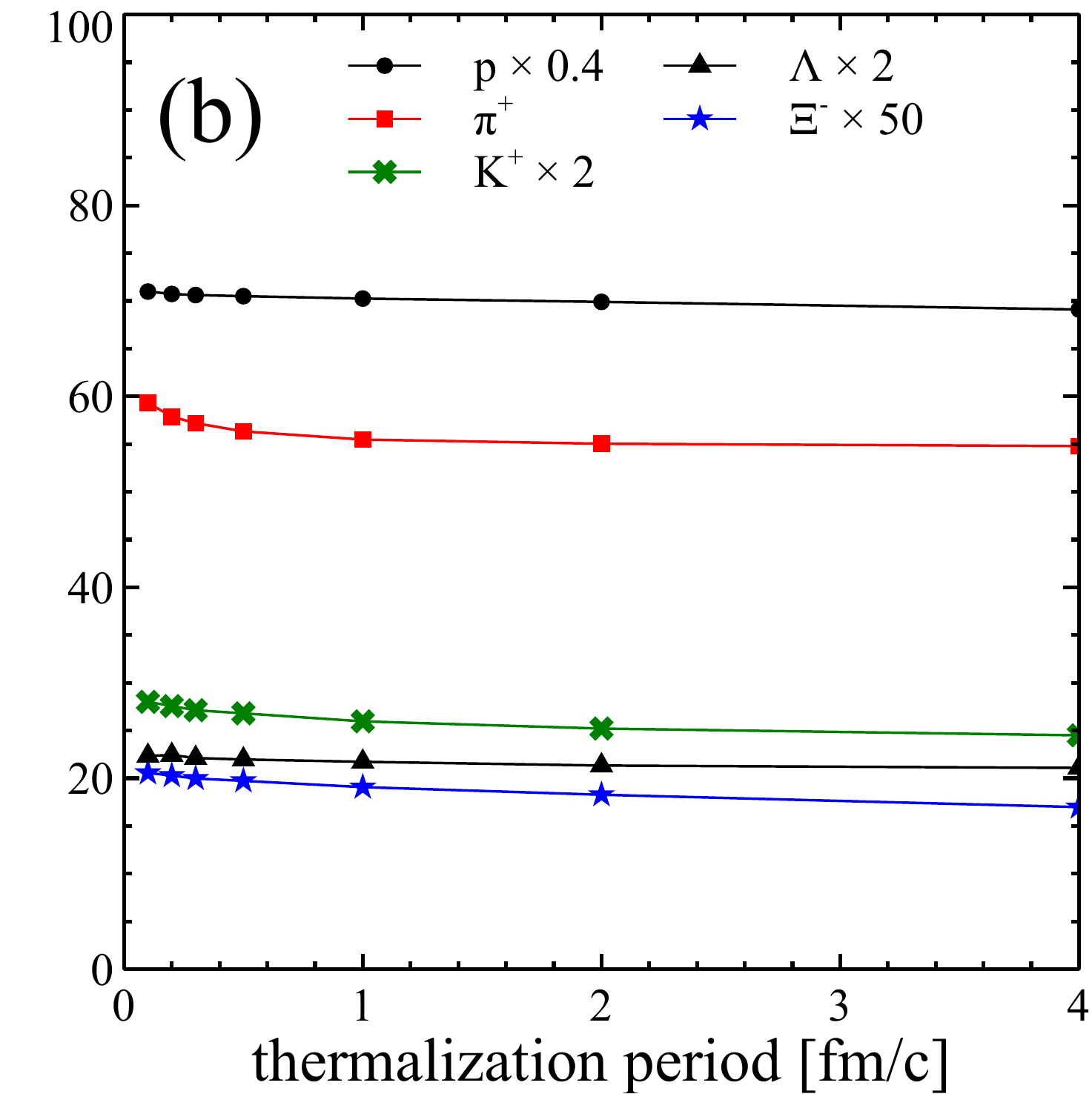} \\ 
  \includegraphics[width=0.49\textwidth]{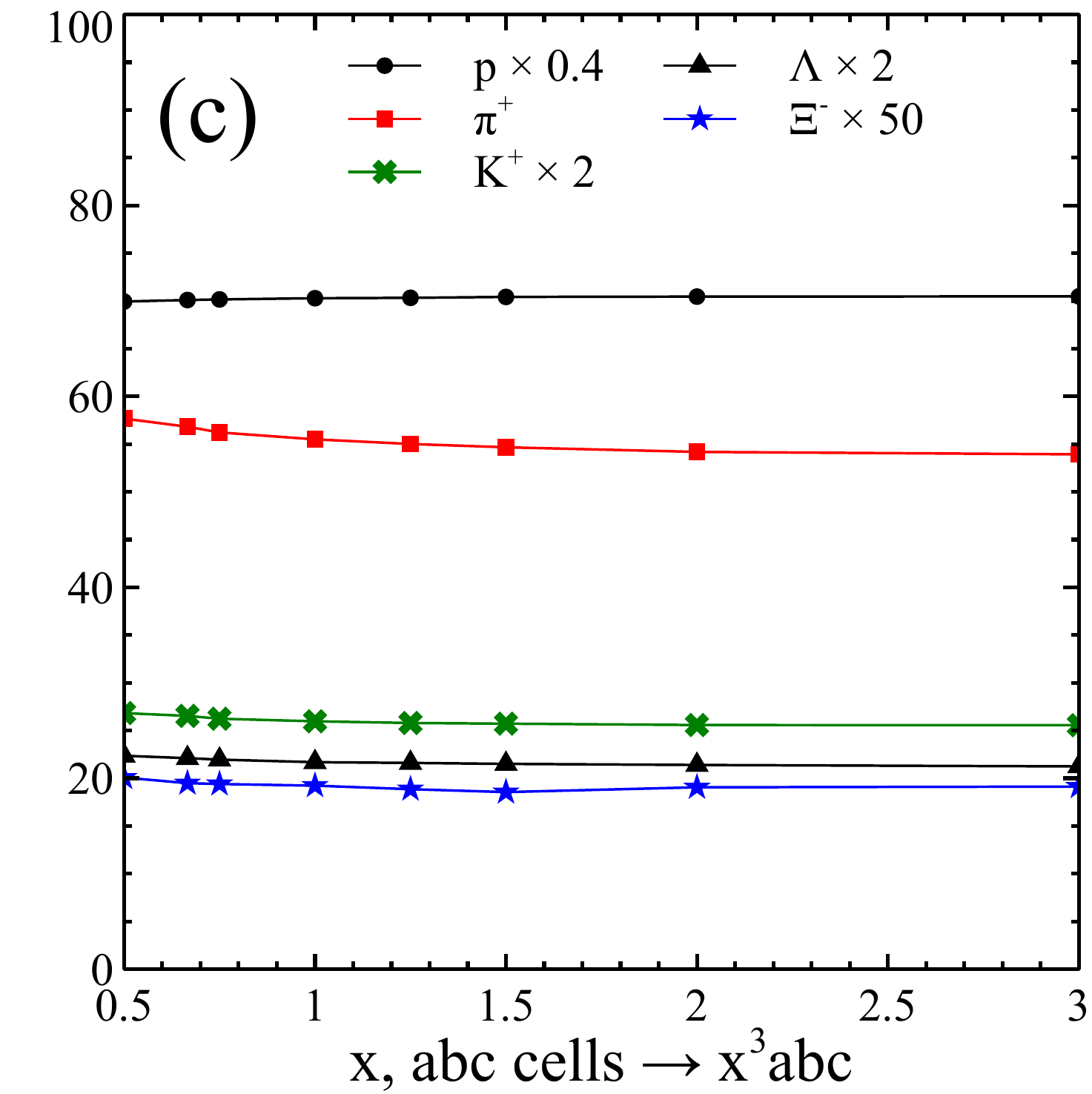}
  \includegraphics[width=0.49\textwidth]{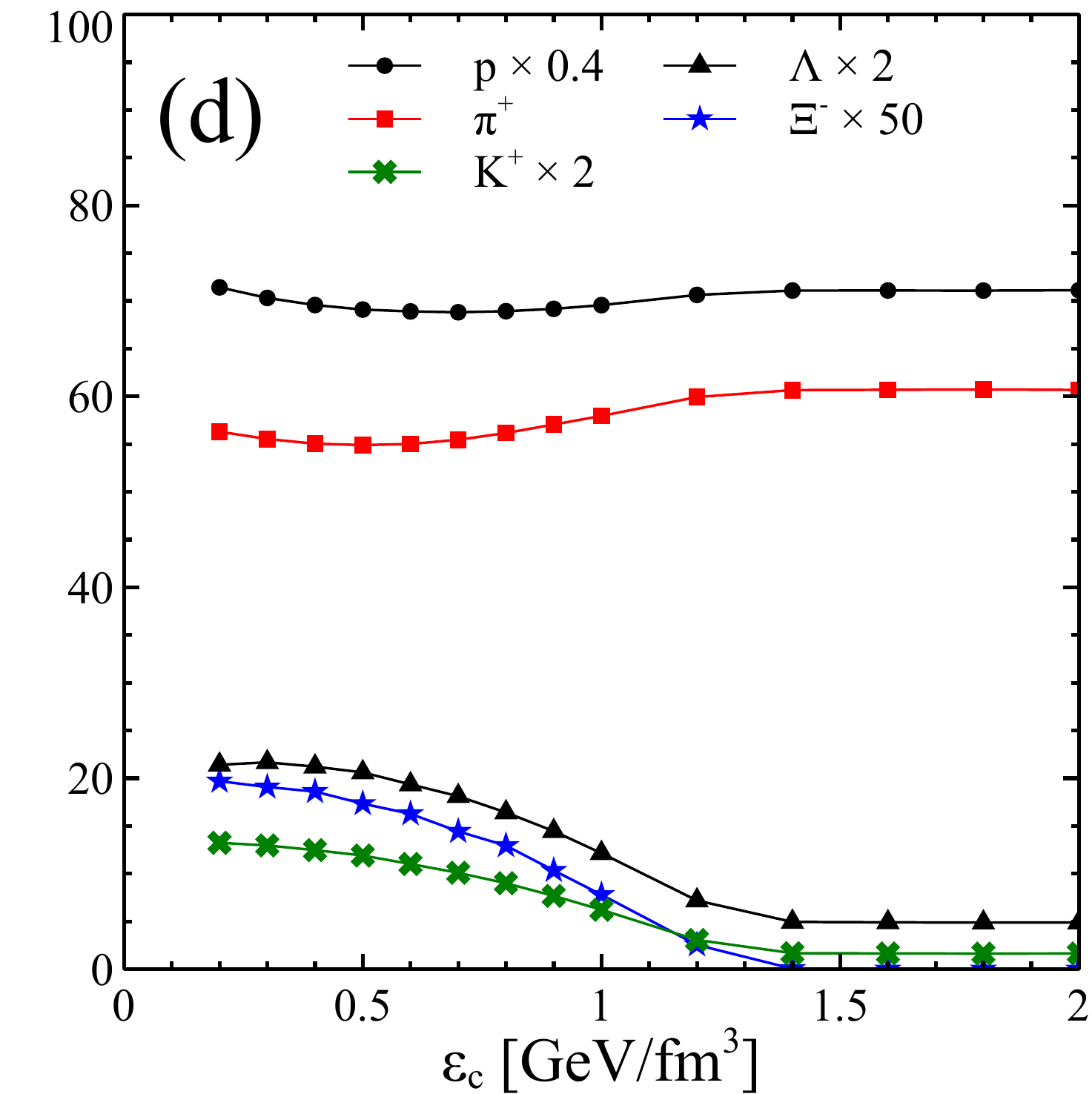}  
  \caption{Central AuAu collision at $\sqrt{s} = 3$ GeV calculated within SMASH with effective treatment of N-particle collisions. Final hadron multiplicities are shown versus test particle number (a), thermalization period $\Delta t_{th}$ (b), grid spacing (panel c, $x$ denotes the factor for number of cells in one dimension, $x = 2$ means that the grid is 8 times denser) and energy density $\epsilon_c$, above which the thermalization is forced (d).}
  \label{Fig:AuAu_param}
\end{figure}

\begin{figure}
  \centering
  \includegraphics[width=0.49\textwidth]{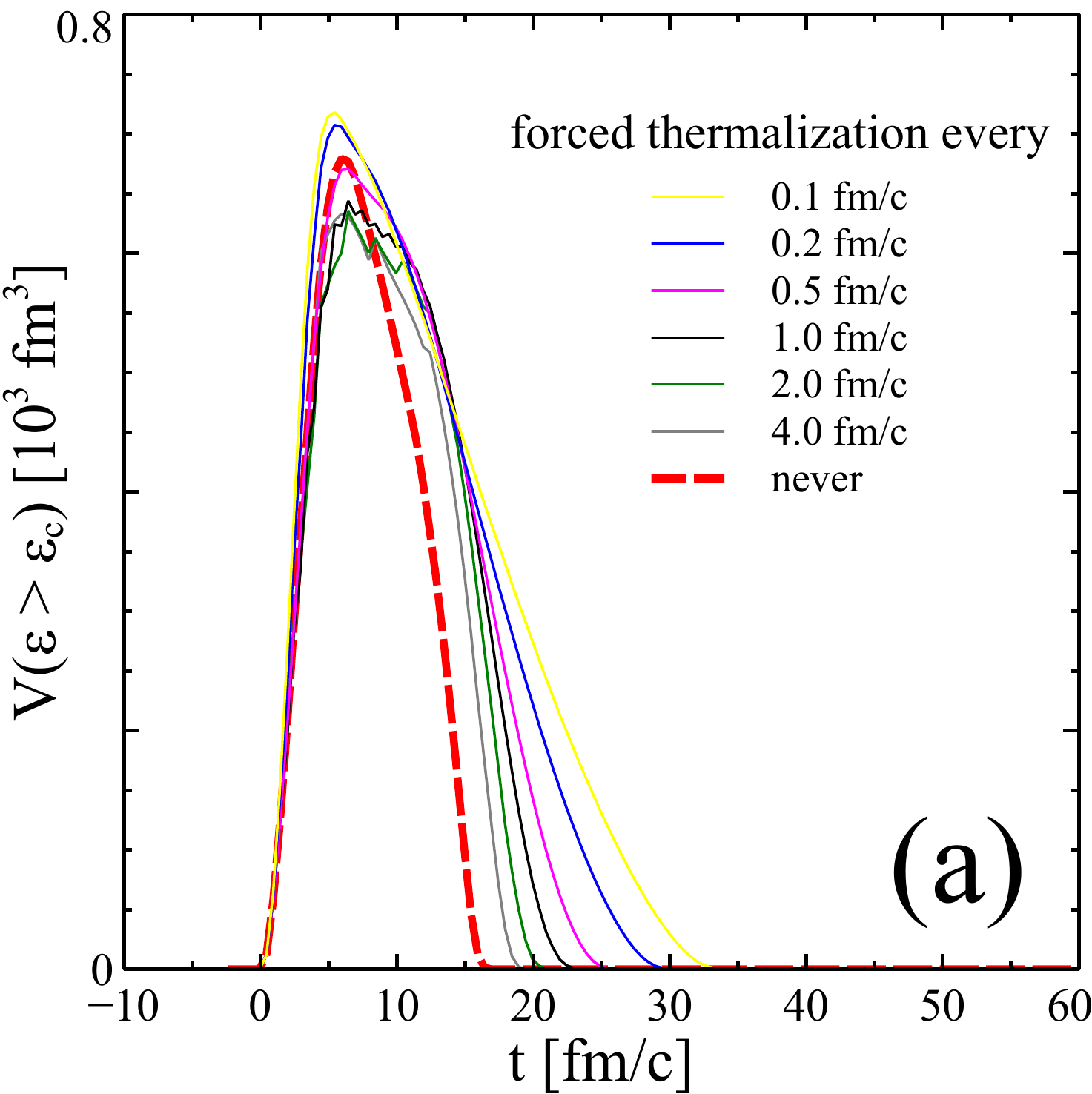}
  \includegraphics[width=0.49\textwidth]{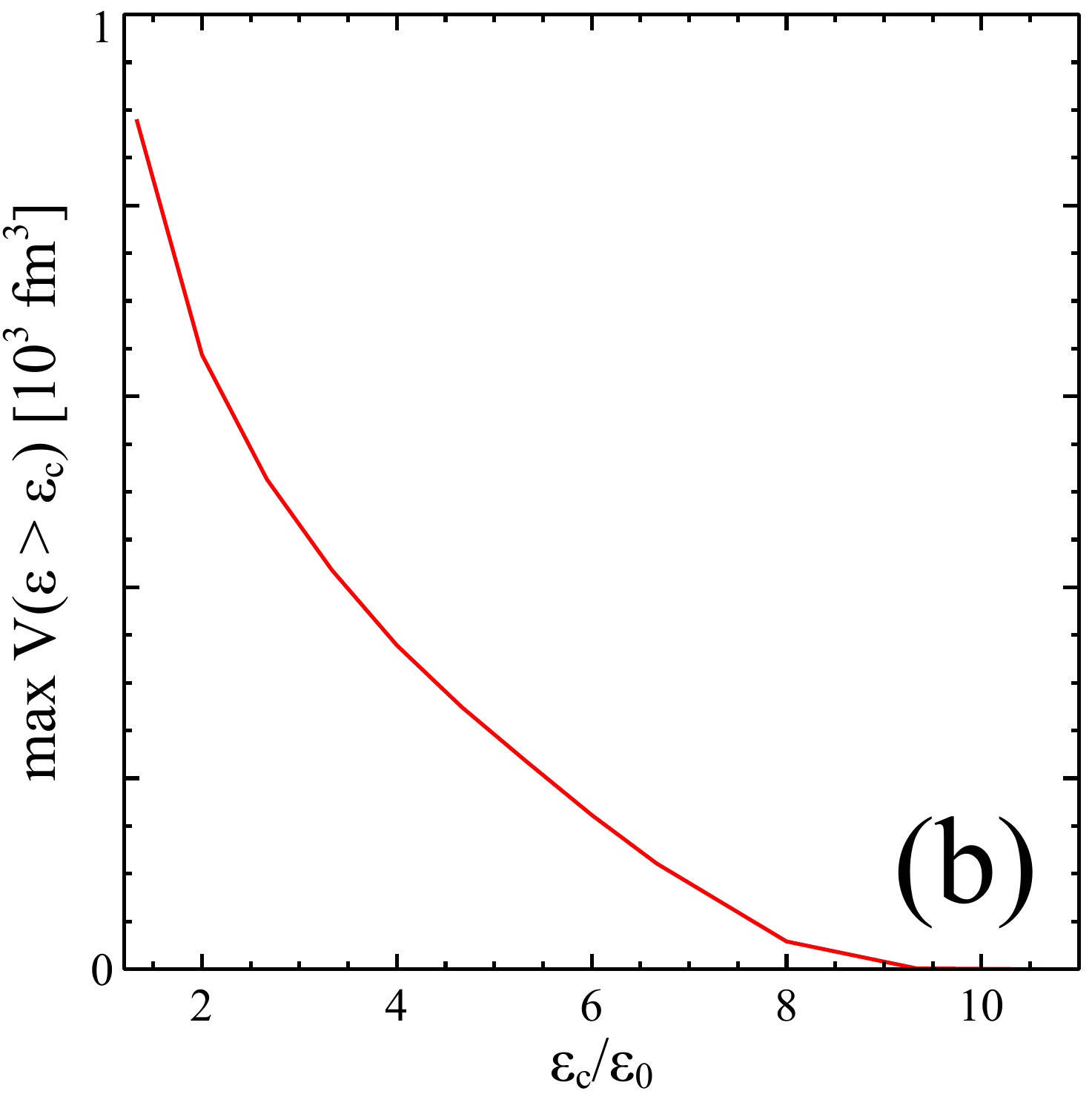}  
  \caption{Volume of thermalization region versus thermalization period $\Delta t_{th}$ (a) and maximal volume versus $\epsilon_c$ (b). Central AuAu collisions at $\sqrt{s} = 3$ GeV simulated by SMASH with effective treatment of N-particle collisions.}
  \label{Fig:AuAu_Vec}
\end{figure}

After studying the effect of forced thermalization in a simple controlled setup, we investigate its implications in heavy ion collisions. To understand our results better, we also consider the influence of the thermalization parameters, such as the thermalization period $\Delta t_{th}$, the thermalization grid spacing and the energy density $\epsilon_c$, above which the thermalization is enforced. In Fig. \ref{Fig:AuAu_param} one can see the effects of varying these parameters. We take $N_{test} = 10$, $\Delta t_{th} = 1$ fm/c, thermalization grid spacing 0.5 fm in the beam direction and 1 fm in the transverse plane and $\epsilon_c = 0.3$ GeV/fm$^3$, and vary these parameters one by one, keeping all the rest constant. As one can see in Fig. \ref{Fig:AuAu_param}, the dependence of the multiplicities on the test particles number saturate at $N_{test} = 10$, which is the reason we chose this number for further investigations. The grid spacing does not affect the final multiplicities, except a small effect on pions. The grid spacing has no physical meaning and ideally results do not depend on it. Surprisingly, $\Delta t_{th}$ also plays a rather small role, even though multiplicities are decreasing with a larger thermalization period, as expected. The dependency on $\epsilon_c$ is also predictable - in the limiting case of high $\epsilon_c$, no thermalization takes place at all, because such high energy densities are never reached in the collision. So for high $\epsilon_c$ SMASH with forced thermalization is equivalent to the normal SMASH cascade. This is also illustrated by Fig. \ref{Fig:AuAu_Vec}. For low $\epsilon_c$, a significant volume is thermalized during the evolution, which drastically increases strange particles multiplicities. This can be attributed to the fact that hadronic interactions do not provide as much strangeness production as a statistical model would predict. In Fig. \ref{Fig:AuAu_Vec} one can also see that the lifetime of the high-density region is prolonged due to the forced thermalization. This is in line with the previously described sphere scenario: transport with forced thermalization becomes closer to the hydrodynamical regime. The observable consequence of such behaviour may be larger HBT radii, compared to pure transport.

\begin{figure}
  \centering
  \includegraphics[width=0.5\textwidth]{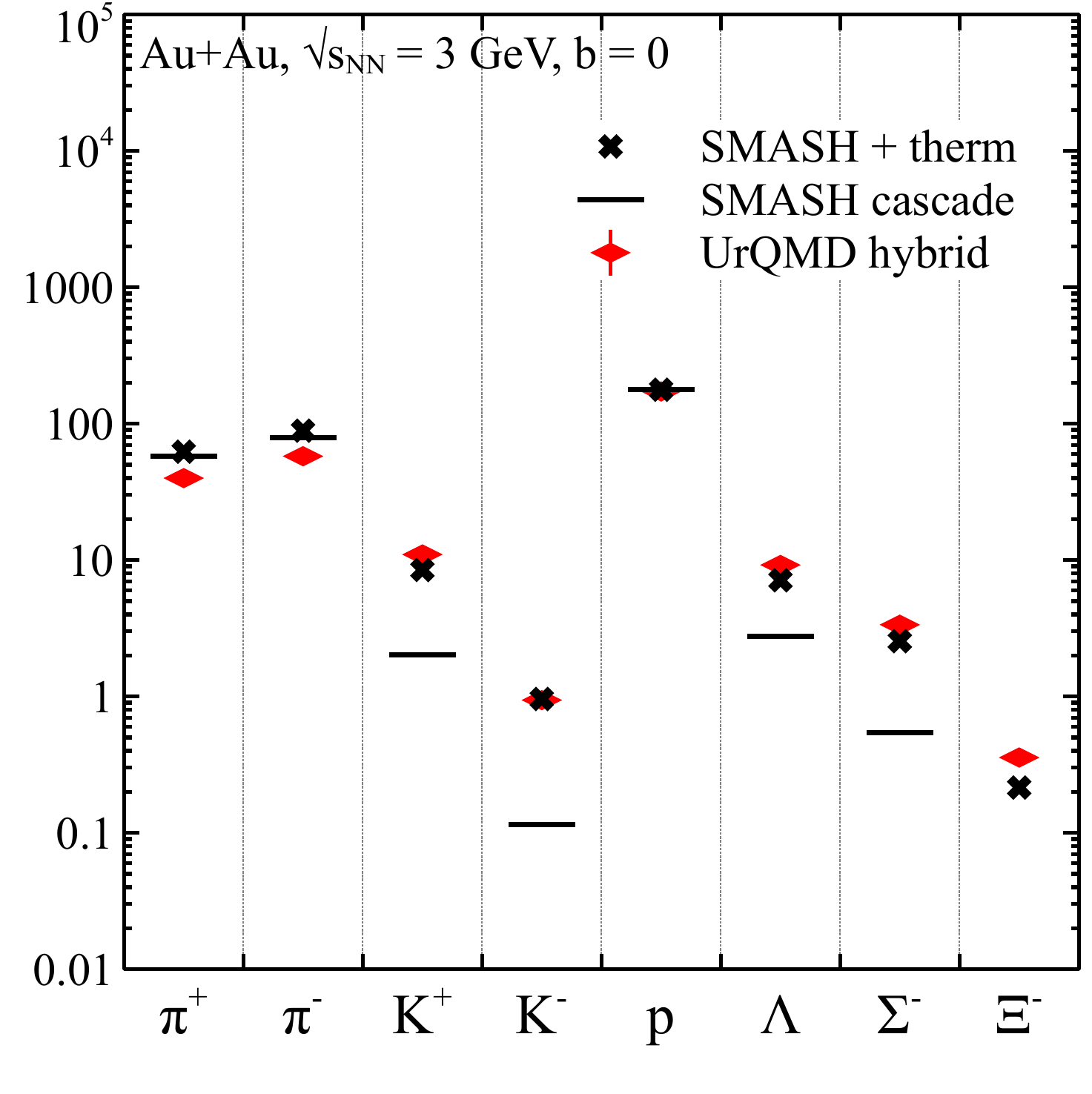}
  \caption{Multiplicities in central AuAu collision at $\sqrt{s} = 3$ GeV are compared for SMASH, SMASH with thermalization, and UrQMD hybrid.}
  \label{Fig:AuAu_smash_urqmd}
\end{figure}

\begin{figure}
  \centering
  \includegraphics[width=0.49\textwidth]{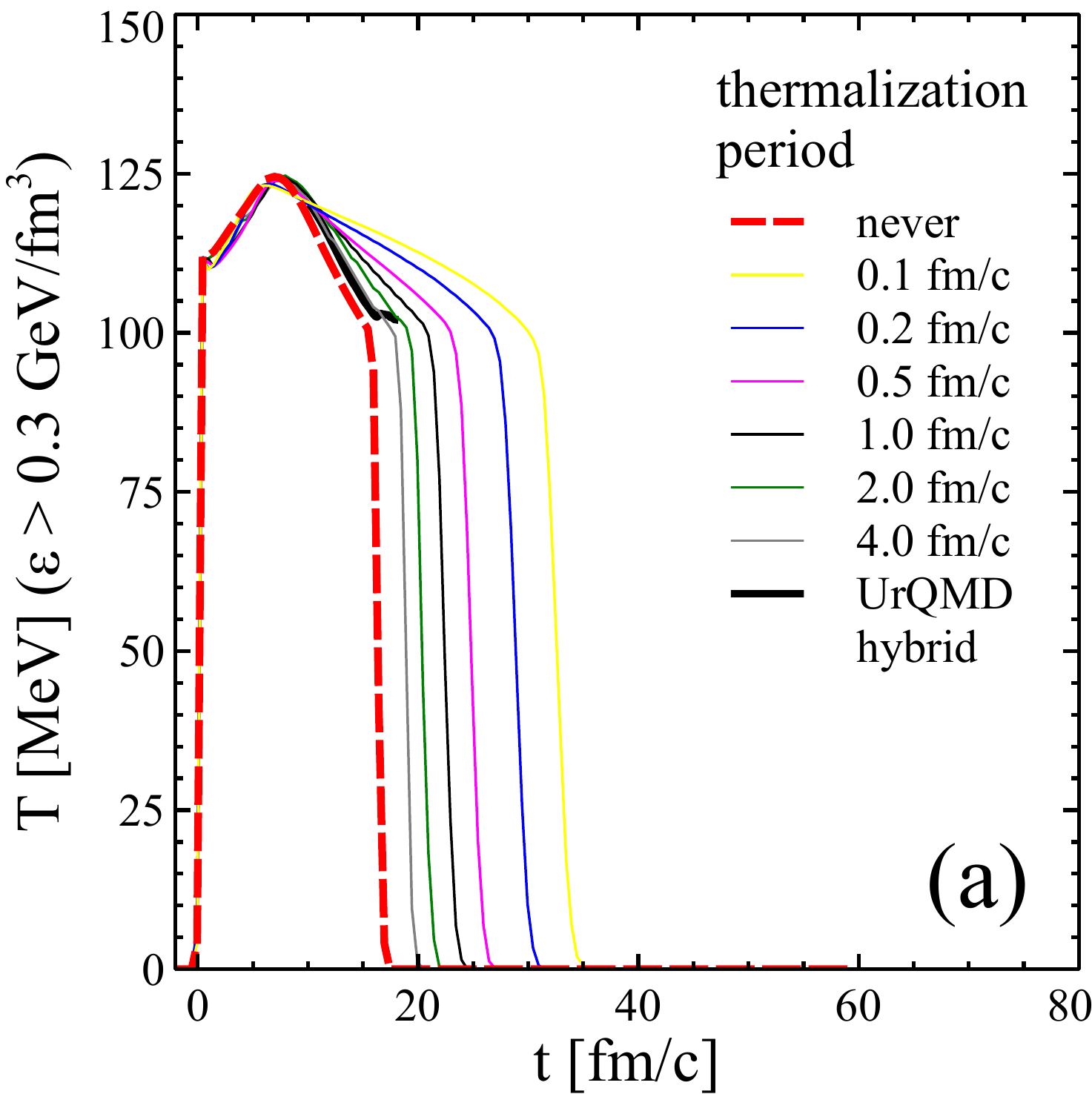}
  \includegraphics[width=0.49\textwidth]{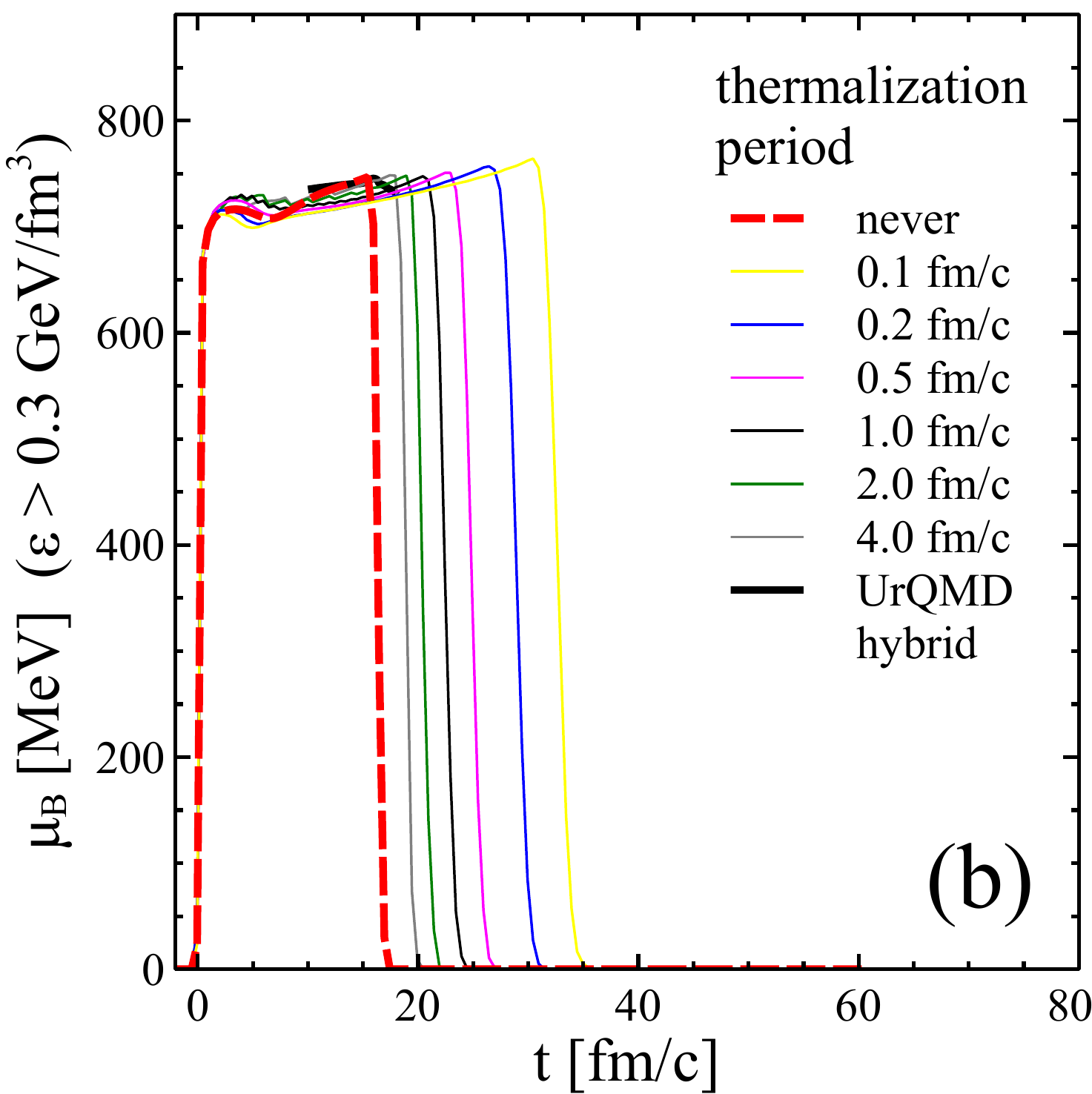}  
  \caption{Average temperature (a) and baryon chemical potential (b) inside of thermalization region for different thermalization periods. Averages are weighted with energy density, i.e. $\langle T \rangle = \sum_r T(r) \epsilon(r) / \sum_r \epsilon(r)$. Central AuAu collisions at $\sqrt{s} = 3$ GeV simulated by SMASH with (solid lines) or without (dashed lines) effective treatment of N-particle collisions. Black solid lines correspond to UrQMD hybrid approach.}
  \label{Fig:AuAu_Tmu}
\end{figure}

Another consequence of our model is a drastic increase of strangeness. This is not surprising, because in the pure transport strange particles are far from thermal equilibration. The effects of our forced thermalization on multiplicities are shown in Fig. \ref{Fig:AuAu_smash_urqmd}, where we compare 3 models: SMASH, SMASH with thermalization, and UrQMD hybrid  \cite{Petersen:2008dd}. The starting time of the thermalization is taken to match that of the hybrid approach. Energy density $\epsilon_c$ is set to $2\epsilon_0$, in the UrQMD hybrid particlization energy density is also set to $2\epsilon_0$. One can see that in terms of multiplicities our model behaves similarly to the UrQMD hybrid approach, even though the underlying transport codes have significant differences in terms of resonance properties and strangeness production. From the Fig. \ref{Fig:AuAu_Tmu} one can see that the average $T$ and $\mu_B$ inside of thermalization/hydrodynamical region are similar in all three approaches, which makes comparison sensible.

\begin{figure}
  \centering
  \includegraphics[width=0.5\textwidth]{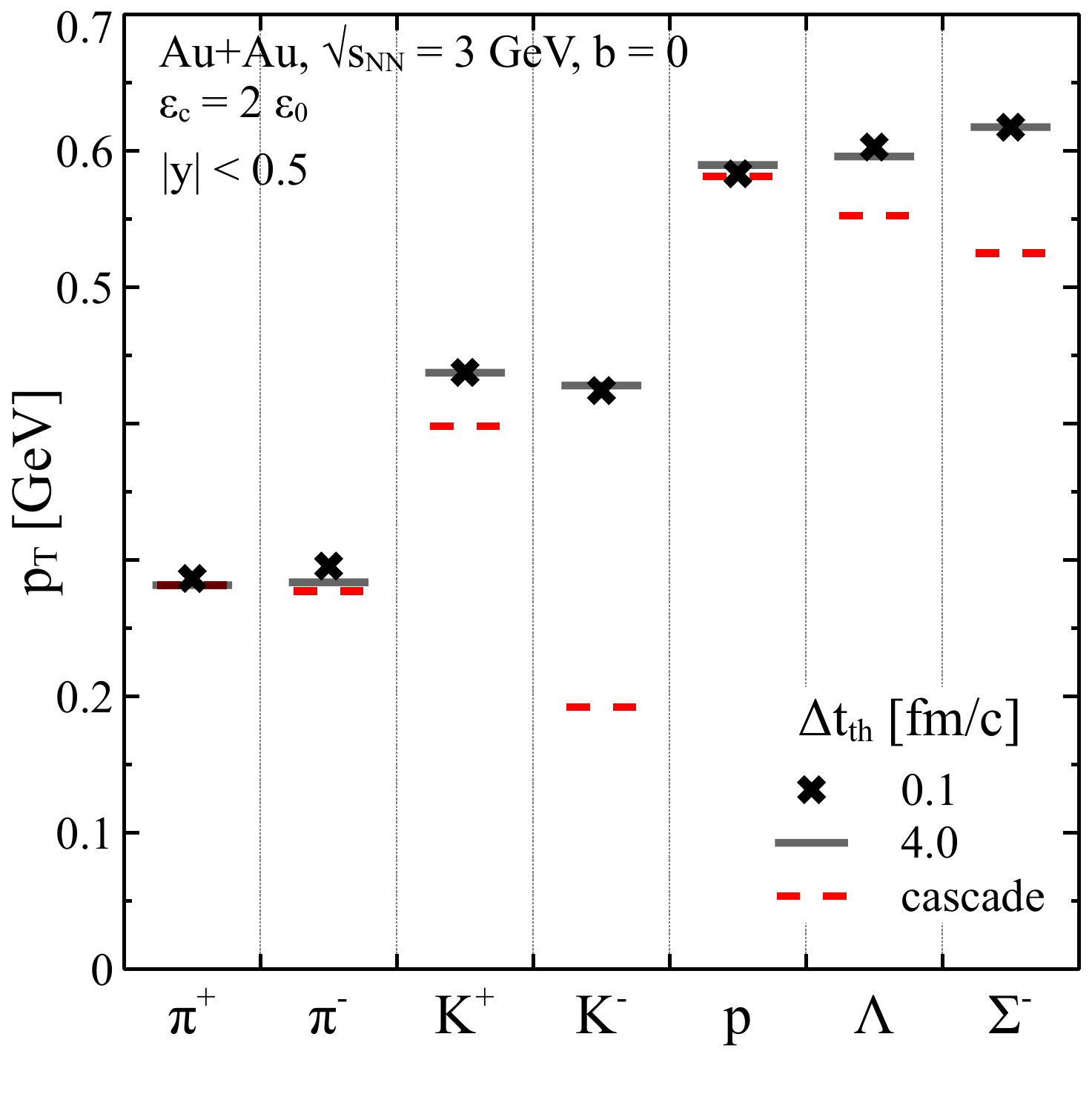}
  \caption{Mean transverse momentum $p_T = \sqrt{p_x^2 + p_y^2}$ at midrapidity in central AuAu collision at $\sqrt{s_{NN}} = 3$ GeV. The model with the forced thermalization is compared to cascade.}
  \label{Fig:AuAu_mean_pt}
\end{figure}

One more consequence of the forced thermalization is that the pressures in the longitudinal and transverse directions rapidly equalize. This means that particles from larger rapidity are redirected to midrapidity and transverse momentum increases. This is illustrated by Fig. \ref{Fig:AuAu_mean_pt}, which shows an increase of mean $p_T$ for all particles. One can see that the mean $p_T$ is insensitive to the thermalization period, but it is quite sensitive to the forced thermalization itself. The most dramatic effect can be seen for $K^-$. We assume that this is because, unlike $K^+$ that can be produced in $NN \to \Lambda K^+$ reactions, more than 80\% of $K^-$ are produced in the secondary strangeness exchange $\pi\Sigma\to N K^-$, $\pi\Lambda\to N K^-$ and $\Sigma^* \to N K^-$. Strangeness exchange reactions preferably deliver their products to high rapidities. Due to the forced thermalization $K^-$ produced at high rapidities are redirected to midrapidity.

\section{Summary and Outlook}

Transport and hydrodynamic approaches are widely used for the dynamical description of heavy ion collisions. Most hadron transport approaches are restricted to binary interactions only and therefore not applicable anymore for the hot and dense stages of the reaction. In this work, we try to overcome this restriction and investigated the effect of incorporating N-particle collisions in a hadronic transport approach in the regions of high energy density. Unlike in hybrid approaches, we did not switch to hydrodynamics to describe the thermalized region, but rather performed thermalization directly in the transport. Also, unlike hybrid approaches, this approach automatically guarantees that the high density and the low density part can exchange particles and that transition hypersurface is determined dynamically. We have implemented and tested this approach using the SMASH hadronic transport approach as a basis. 

First, we have shown, that the biased Becattini-Ferroni algorithm for sampling particles observing conservation laws for all quantum numbers is the most reliable one while being reasonably efficient. In an expanding sphere scenario we have demonstrated that SMASH with effective N-particle collisions exhibits intermediate behaviour between hydrodynamics and transport. The closeness to hydrodynamics can be regulated by the thermalization period - the more often one thermalizes, the closer the result to hydrodynamics. 

Simulating heavy ion collisions with our model we found the following features. Compared to transport, more strangeness is produced, the mean $p_T$ is increased due to pressure isotropization and the high-density region lives longer. All these features are similar to hybrid approaches, but without the above mentioned drawbacks. The final multiplicities are not dependent on grid spacing and only slightly on thermalization period. The values saturate for $N_{\rm test}=10$ and the behaviour as a function of the critical energy density is as expected. Overall, the effective N-particle collisions lead to the expected results and the straight forward tests look promising. 

In the thermalization procedure one needs the EoS to determine local temperature and chemical potentials. For this purpose we have applied a hadron gas EoS, consistent with SMASH hadron content. One can also apply another EoS, for example an EoS with a phase transition, but between thermalizations the propagated degrees of freedom will be still hadrons, which seems inconsistent. If the quark-gluon plasma only exists at high energy densities and at the edges of the thermalized blobs the degrees of freedom are still hadronic, this might allow the direct investigation of EoS of strongly-interacting matter without explicit hydrodynamic evolution. In the future, these studies will be extended to higher collision energies and compared to existing experimental data and provide predictions for upcoming experiments.

\section{Acknowledgements}

The authors thank L.~G. Pang, H.~Niemi and J.~Steinheimer for fruitful discussions, as well as J.-B.~Rose for reading the manuscript. Computational resources have been provided by the Center for Scientific Computing (CSC) at the Goethe-University of Frankfurt. The authors acknowledge funding of a Helmholtz Young Investigator Group VH-NG-822 from the Helmholtz Association and GSI. This work was supported by the Helmholtz International Center for the Facility for Antiproton and Ion Research (HIC for FAIR) within the framework of the Landes-Offensive zur Entwicklung Wissenschaftlich-Ökonomischer Exzellenz (LOEWE) program launched by the State of Hesse. D.~O. acknowledges support by the Deutsche Telekom Stiftung and Helmholtz Graduate School for Hadron and Ion Research (HGS-HIRe). 

\section{References}

\bibliography{inspire,noninspire}

\begin{thebibliography}{10}

\bibitem{Aoki:2006we}
Y.~Aoki, G.~Endrodi, Z.~Fodor, S.~D. Katz, and K.~K. Szabo, ``{The Order of the
  quantum chromodynamics transition predicted by the standard model of particle
  physics},'' {\em Nature}, vol.~443, pp.~675--678, 2006.

\bibitem{Soltz:2015ula}
R.~A. Soltz, C.~DeTar, F.~Karsch, S.~Mukherjee, and P.~Vranas, ``{Lattice QCD
  Thermodynamics with Physical Quark Masses},'' {\em Ann. Rev. Nucl. Part.
  Sci.}, vol.~65, pp.~379--402, 2015.

\bibitem{Huovinen:2001cy}
P.~Huovinen, P.~F. Kolb, U.~W. Heinz, P.~V. Ruuskanen, and S.~A. Voloshin,
  ``{Radial and elliptic flow at RHIC: Further predictions},'' {\em Phys. Lett.
  B}, vol.~503, pp.~58--64, 2001.

\bibitem{Kolb:2003dz}
P.~F. Kolb and U.~W. Heinz, ``{Hydrodynamic description of ultrarelativistic
  heavy ion collisions},'' 2003.

\bibitem{Rischke:1995ir}
D.~H. Rischke, S.~Bernard, and J.~A. Maruhn, ``{Relativistic hydrodynamics for
  heavy ion collisions. 1. General aspects and expansion into vacuum},'' {\em
  Nucl. Phys. A}, vol.~595, pp.~346--382, 1995.

\bibitem{Gale:2013da}
C.~Gale, S.~Jeon, and B.~Schenke, ``{Hydrodynamic Modeling of Heavy-Ion
  Collisions},'' {\em Int. J. Mod. Phys. A}, vol.~28, p.~1340011, 2013.

\bibitem{Molnar:2000jh}
D.~Molnar and M.~Gyulassy, ``{New solutions to covariant nonequilibrium
  dynamics},'' {\em Phys. Rev. C}, vol.~62, p.~054907, 2000.

\bibitem{Xu:2004mz}
Z.~Xu and C.~Greiner, ``{Thermalization of gluons in ultrarelativistic heavy
  ion collisions by including three-body interactions in a parton cascade},''
  {\em Phys. Rev. C}, vol.~71, p.~064901, 2005.

\bibitem{Lin:2004en}
Z.-W. Lin, C.~M. Ko, B.-A. Li, B.~Zhang, and S.~Pal, ``{A Multi-phase transport
  model for relativistic heavy ion collisions},'' {\em Phys. Rev. C}, vol.~72,
  p.~064901, 2005.

\bibitem{Cassing:2009vt}
W.~Cassing and E.~L. Bratkovskaya, ``{Parton-Hadron-String Dynamics: an
  off-shell transport approach for relativistic energies},'' {\em Nucl. Phys.
  A}, vol.~831, pp.~215--242, 2009.

\bibitem{Bass:1998ca}
S.~A. Bass {\em et~al.}, ``{Microscopic models for ultrarelativistic heavy ion
  collisions},'' {\em Prog. Part. Nucl. Phys.}, vol.~41, pp.~255--369, 1998.
\newblock [Prog. Part. Nucl. Phys.41,225(1998)].

\bibitem{Bleicher:1999xi}
M.~Bleicher {\em et~al.}, ``{Relativistic hadron hadron collisions in the
  ultrarelativistic quantum molecular dynamics model},'' {\em J. Phys. G},
  vol.~25, pp.~1859--1896, 1999.

\bibitem{Niemi:2014wta}
H.~Niemi and G.~S. Denicol, ``{How large is the Knudsen number reached in fluid
  dynamical simulations of ultrarelativistic heavy ion collisions?},'' 2014.

\bibitem{Shen:2014vra}
C.~Shen, Z.~Qiu, H.~Song, J.~Bernhard, S.~Bass, and U.~Heinz, ``{The
  iEBE-VISHNU code package for relativistic heavy-ion collisions},'' {\em
  Comput. Phys. Commun.}, vol.~199, pp.~61--85, 2016.

\bibitem{Werner:2010aa}
K.~Werner, I.~Karpenko, T.~Pierog, M.~Bleicher, and K.~Mikhailov,
  ``{Event-by-Event Simulation of the Three-Dimensional Hydrodynamic Evolution
  from Flux Tube Initial Conditions in Ultrarelativistic Heavy Ion
  Collisions},'' {\em Phys. Rev. C}, vol.~82, p.~044904, 2010.

\bibitem{Werner:2012xh}
K.~Werner, I.~Karpenko, M.~Bleicher, T.~Pierog, and S.~Porteboeuf-Houssais,
  ``{Jets, Bulk Matter, and their Interaction in Heavy Ion Collisions at
  Several TeV},'' {\em Phys. Rev. C}, vol.~85, p.~064907, 2012.

\bibitem{Petersen:2010cw}
H.~Petersen, G.-Y. Qin, S.~A. Bass, and B.~Muller, ``{Triangular flow in
  event-by-event ideal hydrodynamics in Au+Au collisions at $\sqrt{s_{\rm
  NN}}=200A$ GeV},'' {\em Phys. Rev. C}, vol.~82, p.~041901, 2010.

\bibitem{Petersen:2011sb}
H.~Petersen, ``{Identified Particle Spectra and Anisotropic Flow in an
  Event-by-Event Hybrid Approach in Pb+Pb collisions at $\sqrt{s_{\rm
  NN}}=2.76$ TeV},'' {\em Phys. Rev. C}, vol.~84, p.~034912, 2011.

\bibitem{Petersen:2014yqa}
H.~Petersen, ``{Anisotropic flow in transport + hydrodynamics hybrid
  approaches},'' {\em J. Phys. G}, vol.~41, no.~12, p.~124005, 2014.

\bibitem{Cooper:1974mv}
F.~Cooper and G.~Frye, ``{Comment on the Single Particle Distribution in the
  Hydrodynamic and Statistical Thermodynamic Models of Multiparticle
  Production},'' {\em Phys. Rev. D}, vol.~10, p.~186, 1974.

\bibitem{Huovinen:2012is}
P.~Huovinen and H.~Petersen, ``{Particlization in hybrid models},'' {\em Eur.
  Phys. J. A}, vol.~48, p.~171, 2012.

\bibitem{Bugaev:1996zq}
K.~A. Bugaev, ``{Shock - like freezeout in relativistic hydrodynamics},'' {\em
  Nucl. Phys. A}, vol.~606, pp.~559--567, 1996.

\bibitem{Bugaev:1999wz}
K.~A. Bugaev and M.~I. Gorenstein, ``{Particle freezeout in selfconsistent
  relativistic hydrodynamics},'' 1999.

\bibitem{Anderlik:1998cb}
C.~Anderlik, Z.~I. Lazar, V.~K. Magas, L.~P. Csernai, H.~Stoecker, and
  W.~Greiner, ``{Nonideal particle distributions from kinetic freezeout
  models},'' {\em Phys. Rev. C}, vol.~59, pp.~388--394, 1999.

\bibitem{Oliinychenko:2014tqa}
D.~Oliinychenko, P.~Huovinen, and H.~Petersen, ``{Systematic Investigation of
  Negative Cooper-Frye Contributions in Heavy Ion Collisions Using
  Coarse-grained Molecular Dynamics},'' {\em Phys. Rev. C}, vol.~91, no.~2,
  p.~024906, 2015.

\bibitem{Cassing:2001ds}
W.~Cassing, ``{Anti-baryon production in hot and dense nuclear matter},'' {\em
  Nucl. Phys. A}, vol.~700, pp.~618--646, 2002.

\bibitem{Pan:2014caa}
Y.~Pan and S.~Pratt, ``{Baryon annihilation and regeneration in heavy ion
  collisions},'' {\em Phys. Rev. C}, vol.~89, no.~4, p.~044911, 2014.

\bibitem{Aguiar:2000hw}
C.~E. Aguiar, T.~Kodama, T.~Osada, and Y.~Hama, ``{Smoothed particle
  hydrodynamics for relativistic heavy ion collisions},'' {\em J. Phys. G},
  vol.~27, pp.~75--94, 2001.

\bibitem{Steinheimer:2011mp}
J.~Steinheimer and M.~Bleicher, ``{Core-corona separation in the UrQMD hybrid
  model},'' {\em Phys. Rev. C}, vol.~84, p.~024905, 2011.

\bibitem{Weil:2016zrk}
J.~Weil {\em et~al.}, ``{Particle production and equilibrium properties within
  a new hadron transport approach for heavy-ion collisions},'' 2016.

\bibitem{Agashe:2014kda}
K.~A. Olive {\em et~al.}, ``{Review of Particle Physics},'' {\em Chin. Phys.
  C}, vol.~38, p.~090001, 2014.

\bibitem{Graef:2014mra}
G.~Graef, J.~Steinheimer, F.~Li, and M.~Bleicher, ``{Deep sub-threshold $\Xi$
  and $\Lambda$ production in nuclear collisions with the UrQMD transport
  model},'' {\em Phys. Rev. C}, vol.~90, p.~064909, 2014.

\bibitem{Cheng:2001dz}
S.~Cheng, S.~Pratt, P.~Csizmadia, Y.~Nara, D.~Molnar, M.~Gyulassy, S.~E. Vance,
  and B.~Zhang, ``{The Effect of finite range interactions in classical
  transport theory},'' {\em Phys. Rev. C}, vol.~65, p.~024901, 2002.

\bibitem{Becattini:2016xct}
F.~Becattini, J.~Steinheimer, R.~Stock, and M.~Bleicher, ``{Hadronization
  conditions in relativistic nuclear collisions and the QCD pseudo-critical
  line},'' 2016.

\bibitem{Oliinychenko:2015lva}
D.~Oliinychenko and H.~Petersen, ``{Deviations of the Energy-Momentum Tensor
  from Equilibrium in the Initial State for Hydrodynamics from Transport
  Approaches},'' {\em Phys. Rev. C}, vol.~93, no.~3, p.~034905, 2016.

\bibitem{Petersen:2008dd}
H.~Petersen, J.~Steinheimer, G.~Burau, M.~Bleicher, and H.~Stocker, ``{A Fully
  Integrated Transport Approach to Heavy Ion Reactions with an Intermediate
  Hydrodynamic Stage},'' {\em Phys. Rev. C}, vol.~78, p.~044901, 2008.

\bibitem{Huovinen:2009yb}
P.~Huovinen and P.~Petreczky, ``{QCD Equation of State and Hadron Resonance
  Gas},'' {\em Nucl. Phys. A}, vol.~837, pp.~26--53, 2010.

\bibitem{Werner:1995mx}
K.~Werner and J.~Aichelin, ``{Microcanonical treatment of hadronizing the quark
  - gluon plasma},'' {\em Phys. Rev. C}, vol.~52, pp.~1584--1603, 1995.

\bibitem{Becattini:2004rq}
F.~Becattini and L.~Ferroni, ``{Statistical hadronization and hadronic
  microcanonical ensemble. 2.},'' {\em Eur. Phys. J. C}, vol.~38, pp.~225--246,
  2004.
\newblock [Erratum: Eur. Phys. J.66,341(2010)].

\bibitem{Yuan2000}
L.~Yuan and J.~D. Kalbfleisch, ``{On the Bessel Distribution and Related
  Problems},'' {\em Annals of the Institute of Statistical Mathematics},
  vol.~52, no.~3, pp.~438--447, 2000.

\bibitem{Devroye2001}
D.~Luc, ``{Simulating Bessel random variables},'' {\em Statistics and
  Probability Letters}, vol.~57, no.~3, pp.~249--257, 2002.

\end{thebibliography}
\bibliographystyle{ieeetr}

\section{Appendix: Sampling Poissonian-distributed numbers with fixed difference}

In the algorithms described above one needs to sample ${N_1}$ and ${N_2}$ such that $w({N_1},{N_2}) \sim \frac{\nu_1^{N_1}}{{N_1}!} \frac{\nu_2^{N_2}}{{N_2}!} \delta({N_1}-{N_2} = N)$, $N > 0$. Let us rewrite it in terms of distribution for $N_2$:
\begin{eqnarray}
w(N_2) = \sum_{N_1=0}^{\infty} w(N_1,N_2) \sim \frac{\nu_1^{N_2+N}}{(N_2+N)!} \frac{\nu_2^{N_2}}{N_2!} \\
w(N_2) = const \frac{(\nu_1 \nu_2)^{N_2}}{N_2!(N+N_2)!}
\end{eqnarray}
Denoting $a = 2\sqrt{\nu_1 \nu_2}$ and normalizing probabilities, one obtains
\begin{equation}
w(N_2) = \frac{a^{2N_2 + N}}{I_N(a) N_2! (N+N_2)!}
\end{equation}

This is the known Bessel distribution. We take recommendations for sampling it from the paper by Yuan and Kalbfleisch \cite{Yuan2000}. Maximal probability for the Bessel distribution is reached for
\begin{equation}
m = \frac{1}{2} \left(\sqrt{a^2 + N^2} - N \right)
\end{equation}
It is suggested by \cite{Yuan2000} that for $m > 6$ the Bessel distribution is very close to the Gaussian distribution, and for $m \leq 6$ probabilities can be computed explicitly and the number can be sampled from a discrete distribution. Moments of $Y \sim Bes(N,a)$ can be computed as 
\begin{eqnarray}
EY = \frac{1}{2} a R_N(a) \\
EY^2 = EY \left( 1 + \frac{1}{2} a R_{N+1}(a)\right) \,,
\end{eqnarray}
Then the mean $\alpha$ and $\sigma$ of the Gaussian are
\begin{eqnarray}
\alpha = EY \\
\sigma = \sqrt{EY^2 - (EY)^2}
\end{eqnarray}
Here $R_N(a) = \frac{I_{N+1}(a)}{I_N(a)} = [\frac{2(N+1)}{a}, \frac{2(N+2)}{a}, \frac{2(N+3)}{a}, \cdots]$, where $[a_1,a_2,a_3,\cdots]$ denotes the continued fraction $\frac{1}{a_1 + \frac{1}{a_2 + \cdots}}$.

An alternative method used by Becattini and Ferroni in \cite{Becattini:2004rq} is to sample two numbers from Poissonian distributions and reject until the difference is the required one. Devroye points out that this method requires $\frac{e^a}{I_{\nu}(a)}$ rejections on average and is thus only acceptable for moderate values of $a$ and $N$ \cite{Devroye2001}. In terms of our purposes, it means that such method works well only for small enough chemical potentials. For completeness we add that in response to the approximate sampling method by Yuan and Kalbfleisch, Devroye has suggested an exact method \cite{Devroye2001}. However, for our purposes the approximate method is sufficient, as Fig. \ref{Fig:BFefix_sampling} demonstrates.

\end{document}